\documentclass[aps,prd,twocolumn,showpacs,floatfix,preprintnumbers,amsfont,amsmath,amssymb,nofootinbib, superscriptaddress]{revtex4-1}

\usepackage[utf8]{inputenc}
\usepackage{amsmath}
\usepackage{physics}
\usepackage{comment}
\usepackage{bm}
\usepackage{siunitx}
\usepackage{graphicx}
\usepackage{natbib}
\usepackage[hidelinks]{hyperref}
\usepackage{mathtools}
\usepackage{xcolor}
\usepackage{ctable}
\usepackage[normalem]{ulem}
\usepackage[caption=false]{subfig}

\newcommand{\rmd}{{\rm d}}

\newcommand{\vphi}{\varphi}
\newcommand{\msun}{{M_\odot}}
\newcommand{\half}{\frac{1}{2}}

\newcommand{\chieff}{\ensuremath{\chi_{\rm eff}}}

\newcommand{\pint}{p_{\rm{int}}}
\newcommand{\pext}{p_{\rm{ext}}}
\newcommand{\vchi}{\vec{\chi}}
\newcommand{\overlap}[2]{\langle \, #1 \, | \, #2 \, \rangle}
\newcommand{\agn}{AGN J124942.3+344929\,}

\newcommand{\nitz}{N\&C\,}

\newcommand{\changed}[1]{#1}

\begin{document}
\title{Mapping the Likelihood of GW190521 with Diverse Mass and Spin Priors}

\author{Seth Olsen}
\email[]{srolsen@princeton.edu}
\affiliation{\mbox{Department of Physics, Princeton University, Princeton, NJ 08540, USA}}
\author{Javier Roulet}
\affiliation{\mbox{Department of Physics, Princeton University, Princeton, NJ 08540, USA}}
\author{Horng Sheng Chia}
\affiliation{\mbox{School of Natural Sciences, Institute for Advanced Study, Princeton, NJ 08540, USA}}
\author{Liang Dai}
\affiliation{\mbox{Department of Physics, University of California, Berkeley, 366 LeConte Hall, Berkeley, CA 94720, USA}}
\author{Tejaswi Venumadhav}
\affiliation{\mbox{Department of Physics, University of California at Santa Barbara, Santa Barbara, CA 93106, USA}}
\affiliation{\mbox{International Centre for Theoretical Sciences, Tata Institute of Fundamental Research, Bangalore 560089, India}}
\author{Barak Zackay}
\affiliation{\mbox{Dept. of Particle Physics \& Astrophysics, Weizmann Institute of Science, Rehovot 76100, Israel}}
\author{Matias Zaldarriaga}
\affiliation{\mbox{School of Natural Sciences, Institute for Advanced Study, Princeton, NJ 08540, USA}}

\interfootnotelinepenalty=10000

\date{September 30, 2021}

\begin{abstract}
  We map the likelihood of GW190521, the heaviest detected binary black hole (BBH) merger, by sampling under different mass and spin priors designed to be uninformative. We find that a source-frame total mass of $\sim$150 $\msun$ is consistently supported, but posteriors in mass ratio and spin depend critically on the choice of priors. We confirm that the likelihood has a multi-modal structure with peaks in regions of mass ratio representing very different astrophysical scenarios. The unequal-mass region ($m_2 / m_1 < 0.3$) has an average likelihood $\sim$e$^6$ times larger than the equal-mass region ($m_2 / m_1 > 0.3$) and a maximum likelihood $\sim$e$^2$ larger. Using ensembles of samples across priors, we examine the implications of qualitatively different BBH sources that fit the data. We find that the equal-mass solution has poorly constrained spins and at least one black hole mass that is difficult to form via stellar collapse due to pair instability. The unequal-mass solution can avoid this mass gap entirely but requires a negative effective spin and a precessing primary. Either of these scenarios is more easily produced by dynamical formation channels than field binary co-evolution. \changed{Drawing representative samples from each region of the likelihood map, we find a sensitive comoving volume-time $\mathcal{O}(10)$ times larger in the mass gap region than the gap-avoiding region. 
  Considering $D_{\rm{com}}^3 \mathcal{L}$ to account for the distance effect, the likelihood of these representative samples still reverses the advantage to favor the gap-avoiding scenario by a factor of $\mathcal{O}(100)$ before including mass and spin priors.} Posteriors are easily driven away from this high-likelihood region by common prior choices meant to be uninformative, making GW190521 parameter inference sensitive to the assumed mass and spin distributions of mergers in the source's astrophysical channel. This may be a generic issue for similarly heavy events given current detector sensitivity and waveform degeneracies.
\end{abstract}

\maketitle

\section{Introduction}

GW190521 was a compact binary coalescence event detected during the first part of the Third Observing Run (O3a) at the Advanced LIGO and Advanced Virgo observatories~\citep{lvc_event_GW190521}. The LIGO--Virgo Collaboration (LVC) reported that the source was a merger of black holes (BHs) with masses $\sim$85 $\msun$ and $\sim$66 $\msun$, and had an effective spin parameter consistent with zero. The coalescence of the inferred source binary produced a remnant BH with a mass of $\sim$142 $\msun$. This classifies the remnant as an intermediate-mass black hole (IMBH), roughly defined by the mass range $10^2$--$10^6 M_\odot$, and it is the heaviest merger yet observed~\cite{lvc_properties_GW190521}.

The large constituent BH masses inferred for GW190521 have sparked tremendous interest since they imply the detection of binary companions within the black hole upper mass gap between $\sim$45 $\msun$ and $\sim$135 $\msun$, which has been theorized based on the physics of pair instability and pulsational pair instability inside stars~\cite{pre_pairinstability_early1964ApJS, Barkat1967PairInstability, another_pairinstability_old1984ApJ, HegerWoosley2002massgap, Woosley2007pairinstabilitySN, Woosley2017ppisn, Farmer2019LowerEdgeBHMassGap, Chen_massgap_collapse_simulations_2014, yoshida_mass_gap_PPI_simulations, pair_instability_mass_loss}. Challenging the simple picture of black hole formation at the evolutionary end point of very massive stars, this has led to the revisiting of binary co-evolution details that could circumvent such mass limits~\citep{ordinary_formation_Belczynski2020, isolated_evolution_dredge-up_mass-gap_CostaMapelli2020, stellar_merger_scenario_CE_redshift_RanzoJiang2020}, particularly for very low metallicity or metal-free progenitors~\citep{Tanikawa2020popIII, Liu2020PopIIINSC, metallicity_effects_mapelli2017, low_metallicity_1gBHs_to85m_shell_interactions_mass_gap_Farrell2021}. Another possibility for generating BHs in the mass gap is through dynamical formation scenarios in dense stellar environments where merger remnants can merge again within a Hubble time, such as active galactic nuclei (AGN) and star clusters \cite{imbh_from_globular_clusters2002, hierarchical_7merger_scenario2020b, hierarchical_evidence_inGWTC2_assume_dynamical_Kimball2020b, hierarchical_from_dynamical_in_any_star_cluster2020b, hierarchical_from_triples2021, young_star_clusters_heavy_remnants2021, young_star_clusters_populating_mass_gap2020a, agn_accretion_disk_merger_population2020a, mass_gap_agn_bbh_mergers2021, agn_bbh_population_chieff_q_simulation_mckernan_ford2019, bbh_evolution_agn_merger_timescale_ishibashi2020a, hierarchical_mergers_agn_kocsis2019}.

Formation models and population simulations inform one of the most important considerations in Bayesian inference of gravitational wave (GW) source parameters: the assumed prior distribution. While modeled populations can give us analytic and numerical intuition about the astrophysical population of BBH mergers, they too depend on speculative choices for prior distributions. For instance, priors on the metallicity and spin distributions will determine the importance of the first-generation (1g) mass gap BH abundance \cite{Mapelli2020rotationEffectOnMassGap}, and the dynamical formation contribution of second-generation (2g) and higher BHs in the mass gap is sensitive to the priors on the mass and spin distributions \cite{hierarchical_mergerFamily_dynamical_mass_dist_matters2021, dynamical_rates_sensitive_to_priors_Rodriguez_2018}.

Observing the first BBH merger like GW190521 is exciting but comes with the caveat that both the source parameters (namely, mass ratio and spin) and the astrophysical implications of such a system (e.g., violation of a mass gap and population statistics) are sensitive to uncertain priors. This makes it important to identify the conclusions that are robust to reasonable changes in prior assumptions. Given the event's many astrophysical implications, we search for robust features of GW190521 and find that different choices of uninformative prior lead to interpretations with notably different impact on our preliminary statistical understanding of BBH spin-orbit misalignment and the mass gap.

Population analysis has been used to search for a mass gap empirically \cite{mass_gap_from_population2021, poking_holes_mass_gap_from_pop_farr2021}, and GW190521 plays a key role in informing this discussion. Population analyses usually rely on posterior samples to map the likelihood of individual event parameters, which entails the key assumption that the samples represent unbiased draws from the posterior under some fiducial prior that can be reliably mapped to other priors. In agreement with other works \cite{nitz2020_game_over, new_scoop_waveform_model_comparison_xphm2021}, we find that GW190521 is an example for which standard choices of priors cause posterior samplers to under-sample regions of parameter space with high-likelihood solutions, or miss them entirely. The LVC event catalog results are obtained under a uniform detector-frame constituent mass prior and an isotropic spin prior \cite{lvc_GWTC-1, lvc_GWTC-2}, and inferred source parameters can be sensitive to the choice of priors \cite{impact_of_priors_Vitale2017}. One constraining feature of the LVC priors is the strong preference for zero effective spin. An alternative is to adopt the uniform effective spin prior introduced by \citet{flat_chieff_prior_o1event1216_formation_channels2019}, so we compare the these two approaches for GW190521.

The mass prior plays an even larger role than the spin prior in the case of GW190521. The LVC priors led to the inference of a nearly equal-mass binary with both BHs in the mass gap \cite{lvc_properties_GW190521}. In a reweighting of the LVC posterior samples using a population-informed mass prior for the secondary BH, \citet{gap_stradler_fishbach2020} argued the possibility that a mass ratio farther from unity could describe the event and move one or both of the BHs outside the mass gap. Sampling under a mass prior uniform in source-frame total mass and inverse mass ratio, \citet{nitz2020_game_over} (also referred to hereafter as N\&C) found new regions of high likelihood for unequal-mass solutions. \nitz used the IMRPhenomXPHM waveform model \cite{xphm_paper}, and the latest updates to the precession implementation have reduced the advantage in likelihood of the most extreme mass ratio peak \cite{new_scoop_waveform_model_comparison_xphm2021}, but the recent detection of the $(\ell, |m|, n) = (3, 3, 0)$ quasinormal mode in the GW190521 ringdown by \citet{quasinormal_ringdown_PEcorrection_nitzCapano2021} supports the interpretation of an unequal-mass merger.

In this paper we explore the differences between posteriors sampled under the mass priors of \nitz and LVC, paired with the isotropic spin prior and the uniform effective spin prior. We find that GW190521 is a particularly difficult event to analyze due to the signal's small number of cycles in the sensitive band of the detectors. Waveform model degeneracies in the relevant regions of intrinsic parameter space become even harder to resolve, making it difficult for posterior samplers to explore high-likelihood regions where the prior volume is low. We find that the source-frame total mass is robust but the inferred mass ratio and spin are sensitive to the choice of priors. The mass prior used by \nitz favors unequal-mass sources compared to the LVC mass prior's preference for inverse mass ratio near unity, and the likelihood has peaks in both regions. The solution with comparable masses in the mass gap has lower likelihood but an advantage in geometric prior volume, and the solution with a secondary BH below the mass gap strongly favors a primary mass above 100 $\msun$, which would represent the first observation of a merging IMBH.

We find that regardless of spin prior, the unequal-mass solution is characterized by negative effective spin. The equal-mass solution is consistent with zero effective spin, but even for comparable masses the likelihood favors effective spins near $\pm0.3$. When considering primary spin components in the plane of the orbit, the equal-mass solution shows little evidence of precession, but the unequal-mass solution has a drastic preference for nonzero in-plane spin. Thus we find qualitatively different sources in the likelihood peaks, and the mass and spin priors determine which peaks are explored by posterior samplers. Whether we have observed two BHs in the mass gap or a stellar-mass BH orbiting a precessing IMBH with negative effective spin, the system may be tricky to produce with field binary co-evolution, whereas dynamical formation channels can more easily accommodate each interpretation.

This paper is organized as follows: in Section~\ref{sec:priors} we describe the practical considerations that go into posterior sampling and we define our four sampling priors. In Section~\ref{sec:posteriors} we present the posteriors resulting from parameter estimation (PE) under each set of mass and spin priors, illustrating the multi-modal structure of the likelihood. In Section~\ref{sec:likelihood} we map the likelihood manifold to better classify the various scenarios for the source masses and spins. In Section~\ref{sec:discussion} we summarize these results and conclude with a brief discussion of the challenges that GW190521 poses for BBH formation channel models.

\section{Parameter Estimation Priors}\label{sec:priors}

\subsection{Bayesian Inference for GW Sources}

Bayesian parameter estimation of a GW signal seeks to infer the source's intrinsic and extrinsic parameters from the posterior probability distribution
\begin{equation}
    P(\, \pint, \pext \, | \, d \,) \, = \, P( \, d \, | \, \pint, \pext \,) \frac{\Pi(\, \pint, \pext \,)}{\Pi(\, d \,)},
\end{equation}
where $d$ is the strain data, $\pint$ represents the intrinsic source parameters, and $\pext$ represents the extrinsic parameters. $P(\, d \, | \, \pint, \pext \,) \equiv \mathcal L (\pint, \pext; d)$ is the likelihood and $\Pi(\pint, \pext)$ is the prior probability we assign to the parameters. The prior on the data, $\Pi(d)$, along with any other parameter-independent overall factors, can be absorbed into a normalization constant that we will leave out of the notation hereafter.

For quasi-circular (i.e., negligible orbital eccentricity) BBH mergers, where the tidal deformability parameters of Kerr black holes are zero in the adiabatic limit~\cite{BH_zero_tidal_deformability_chia2020tidal, Charalambous:2021mea}, we consider the 8-dimensional intrinsic parameter space spanned by the constituent BH masses and dimensionless spins, $\pint = (m_1, m_2, \vec{\chi}_1, \vec{\chi}_2$), where the mass ordering $m_1 \geq m_2$ determines which BH is labeled the ``primary" versus ``secondary" constituent. The extrinsic parameters constitute the 7-dimensional space of sky location (right ascension and declination, $\alpha$ and $\delta$), luminosity distance ($D_L$), orbital orientation with respect to the line of sight (LOS) at some reference epoch (polar and azimuthal angles of the LOS in the source frame, $\iota$ and $\vphi$, termed inclination and orbital phase), the angular degree of freedom about the LOS ($\psi$, termed polarization phase), and the time of merger ($t_c$).

\subsection{Posterior Sampling Liabilities}

Monte Carlo (MC) integration of the Bayesian evidence is used to sample the posterior distribution over the full 15-dimensional space of parameters. Due to high dimensionality and waveform degeneracies, exploring all the important regions of parameter space can be difficult for sampling algorithms. Sampling in well-measured coordinates with minimal correlation helps reduce the risk of pathological convergence, but it is still possible for likelihood peaks to be missed in regions of low prior volume.

The prior distribution used in sampling is an analytically tractable problem that may be numerically delicate in practice. If we know the true prior distribution of a parameter, then it makes sense to impose that prior during sampling, but this is not the case for intrinsic parameters. The true mass and spin priors depend on physical and statistical characteristics of BBH formation channels. Without knowledge of the true distribution, the goal is to choose priors to be as uninformative as possible (see Section~\ref{sec:comparison} below) so that the posterior is determined by the likelihood.

When we model the LVC detector noise as a stationary Gaussian process, the likelihood's dependence on the source parameters is given by 
\begin{equation}
    \mathcal{L}(p; d) \, \propto \, {\rm e}^{\overlap{d}{h(p)} - \half \overlap{h(p)}{h(p)}} \, ,
\end{equation}
where $h(p)$ is the modeled gravitational waveform corresponding to merger parameters $p \equiv (\pint, \pext)$ and $\overlap{d}{h(p)}$ is its matched-filter overlap with the data (i.e., variance-weighted inner product). If we can find a \textit{sampling prior} (more commonly referred to as a \textit{false prior}) that allows us to accurately sample this likelihood manifold, then different choices of \textit{target prior} can be analyzed in post-processing by reweighting the samples with methods like importance sampling \cite{reweighting_importance_sampling_thrane2019} or prior swapping \cite{prior_swapping_neiswanger2017}.

Reweighting is analytically equivalent to sampling in the target prior as long as the sampling prior is nonzero everywhere that the target prior has support. This does not always hold in practice due to the finite number of samples and the limitations of MC algorithms. Sampling bias may be an important issue for a broader class of events including GW151226 and GW150914 \cite{pe_reliabilityO1_kulkarni_capano2021}, and in agreement with previous studies \cite{nitz2020_game_over, new_scoop_waveform_model_comparison_xphm2021} we find that the standard LVC priors lead to an incomplete map of the GW190521 likelihood manifold. In contrast to previous approaches, we do not assume that any one choice of prior produces samples that can be reweighted reliably. Instead we assume only that sampling GW190521 posteriors under a given prior produces a reliable map of the likelihood in regions which are not suppressed by that sampling prior.

Relaxing the reweighting assumption in favor of this milder likelihood mapping assumption means that we can tolerate the possibility of sampling bias, but we cannot put faith in the posteriors beyond the likelihood map. Despite the lack of an unbiased Bayesian interpretation in this framework, posterior samples can still illustrate the interplay between priors and likelihood. An important lesson from this illustration is that there is no such thing as a truly uninformative prior. Evaluating the information introduced by one prior requires comparison with other priors. Posterior features that are consistent across a range of priors can be thought of as robust to the information introduced by the priors being varied, but features that change are evidently sensitive to that information. Here we sample the GW190521 source parameter space under a physically diverse range of attempts at designing uninformative mass and spin priors.

\subsection{Comparing Uninformative Mass and Spin Priors} \label{sec:comparison}

One attempt to be uninformative is to say that we do not have any prior knowledge about the direction of constituent BH spins, and therefore a natural choice would be to draw spins from independent isotropic distributions. Following similar logic, our lack of prior knowledge about the mass distributions of BBH populations means that one reasonable attempt to be uninformative is to draw the constituent BH detector-frame masses from independent distributions that are uniform over some large range. These are the priors used in the LVC analysis of GW190521 \cite{lvc_properties_GW190521}, and in their source parameter inference throughout the event catalogs \cite{lvc_GWTC-1, lvc_GWTC-2}.

Another attempt to minimize the amount of information introduced by a sampling prior is to make it uniform in the best-measured parameters. For a quasi-circular BBH, a well-measured spin parameter is typically the effective spin, defined as
\begin{equation}
    \chieff \, = \, \frac{1}{1 + q} \big(\vchi_1 \,+\, q \, \vchi_2 \big) \cdot \hat{L}
\end{equation}
where $q = m_2 / m_1 \in (0, 1]$ is the mass ratio and $\hat{L}$ is the direction of the binary's orbital angular momentum. \changed{Just as coupling the constituent masses through the chirp mass gives us a coordinate which is better suited to the space of GW signals than the more physically intuitive variables composing it, so too does coupling the mass ratio and orbit-aligned spin components through the effective spin.} A spin prior that is uniform in $\chieff$ randomizes sampling of the combination of mass ratio and spins that the data is best at describing, allowing the likelihood to more directly determine the posterior distribution. This comes at the cost of disfavoring values of the constituent spin magnitudes close to zero. On the other hand, the isotropic prior is uniform in constituent spin magnitude but suppresses effective spins far from zero. Comparing these approaches gives a better understanding of the information that each choice introduces into posteriors, and the combined ensemble of samples is a more reliable map of the likelihood over the full range of both effective spin and spin magnitude.

For the mass prior, we compare the LVC approach and that of \citet{nitz2020_game_over}, who sample with a prior that is uniform in source-frame total mass, $M \, = \, m_1 \, + \, m_2 \in [80, 300] \, \msun$, and inverse mass ratio, $q^{-1} = m_1 / m_2 \in [1, 25]$. This is a good way to be agnostic in the case of GW190521 because source-frame total mass is the best-measured mass parameter, and randomizing the sampling of inverse mass ratio avoids suppressing unequal-mass (i.e., small mass ratio $\iff$ large inverse mass ratio) regions of parameter space that contain high-likelihood solutions. If we think in terms of prior volume by mass ratio, uniform in $q^{-1}$ favors unequal-mass solutions, whereas the LVC mass prior is better at exploring equal-mass regions of the likelihood manifold. In Section~\ref{sec:posteriors} we see the effect that each choice has on GW190521 posteriors in combination with each spin prior, and in Section~\ref{sec:likelihood} we use the combined ensemble of samples to obtain a more reliable map of the likelihood throughout the mass ratio sampling range.

Hereafter we refer to the prior that is uniform in effective spin as the ``FLAT" spin prior (described in more detail in \cite{flat_chieff_prior_o1event1216_formation_channels2019}). The LVC prior with isotropic independent constituent spins will be called the ``ISO" spin prior. The prior that is uniform in the source-frame total mass and inverse mass ratio will be called the ``NC" mass prior, using the ranges above (following \citet{nitz2020_game_over}). The prior that is uniform in detector-frame constituent masses will be called the ``LVC" mass prior. We extend the LVC mass prior beyond the \changed{bounds from the original publication \cite{lvc_properties_GW190521}} in order to include the entire range of constituent masses where the \nitz posteriors had support.

\subsection{Geometric Extrinsic Priors and Physical Assumptions}

For the extrinsic parameters we use geometric priors. That is, our priors are isotropic in all the BBH and detector orientation angles. For the LVC mass prior we use a distance distribution that is uniform in luminosity volume, following \citet{lvc_properties_GW190521}. For the NC mass prior we use a distance distribution that is uniform in comoving spatial volume, following \citet{nitz2020_game_over}. The comoving and luminosity distances are proportional by a cosmological redshift factor, $D_L / D_{\rm{com}} = 1 + z$, and we compute the redshift using a $\rm{\Lambda C D M}$ cosmology with the Planck 2015 measurements \cite{cosmology_planck2015}. Note that this comoving volume differs from the comoving volume-time ($VT$), which has an additional redshift factor for time dilation.

Neither comoving nor luminosity volume would be exactly correct to hold uniform under the true astrophysical prior. In any situation where the two choices give different answers, we also need to incorporate the redshift evolution of merger rates to construct an unbiased astrophysical distance prior. We have verified that GW190521 is a case in which the difference between posteriors under priors that are uniform in comoving versus luminosity volume is a negligible perturbation on top of the variation between uninformative mass and spin priors. In all cases our sampling region's total spatial volume maps to the luminosity distance range $D_L \leq 10$ Gpc. The prior on coalescence time is uniform over 100 ms centered about GPS time 1242442967.44 at the Livingston (L1) detector.

We assume a quasi-circular inspiral, which is valid as long as the BBH circularization timescale is much smaller than the merger time scale \cite{bbh_circularization_peters1963PhysRev}. This holds in most cases, although dynamical scenarios can produce eccentric mergers under the right conditions (e.g., head-on collision or close-capture, possibly aided by interactions with additional compact companions). Thus another step toward agnostic priors is to allow for nonzero orbital eccentricity. There are currently not enough NR simulations in the space of eccentric BBH mergers to make a reliable surrogate model for parameter estimation with generic spin and mass ratio, so we work in the quasi-circular approximation.

\begin{figure}
    \centering
    \includegraphics[width=\linewidth]{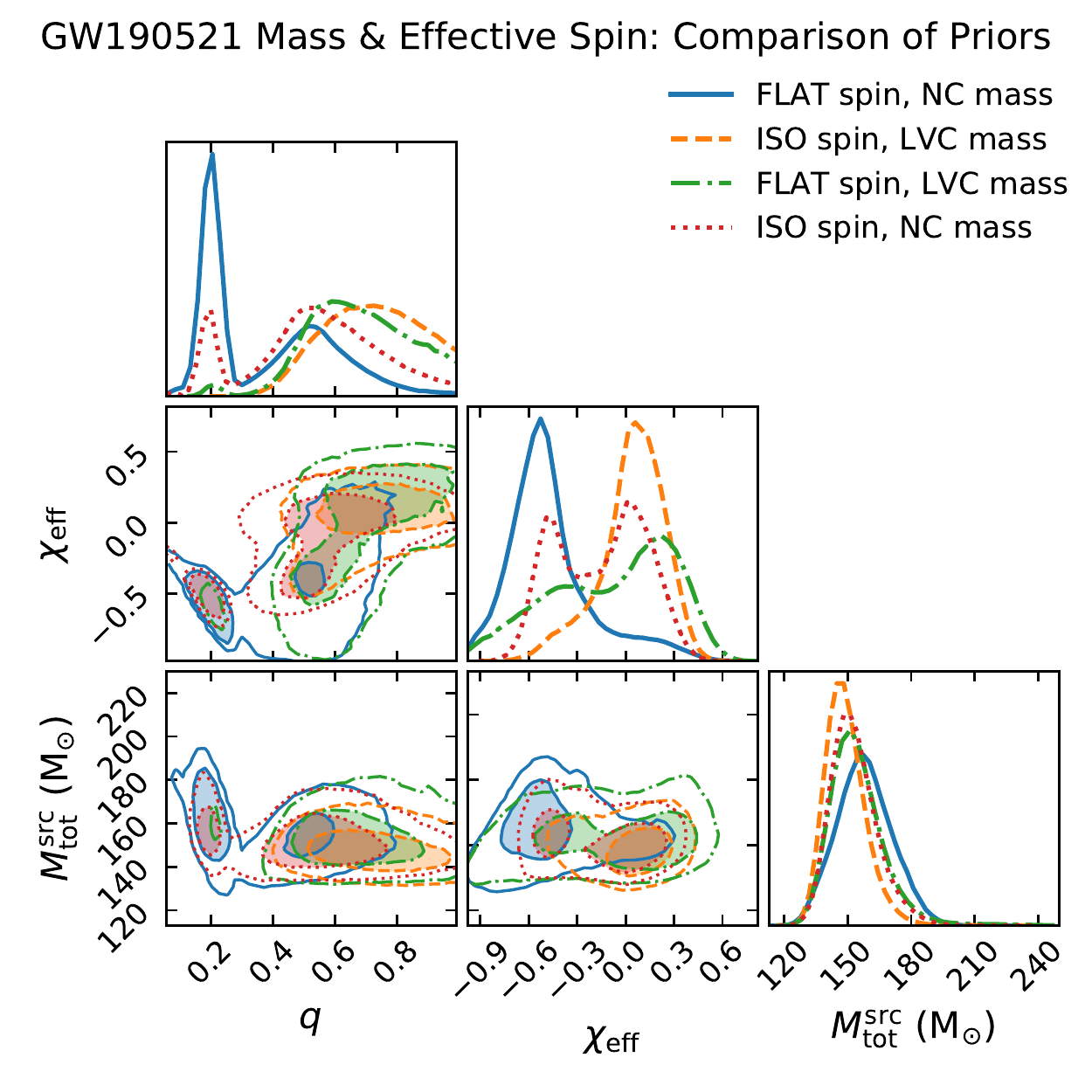}
    \caption{Effective spin and mass posteriors compared across priors. In the 2-dimensional plots, the contours mark 50\% and 90\% confidence intervals. While the source-frame total mass is consistent, $q$ and $\chieff$ are sensitive to choices of priors. Intrinsic parameter regions favored by one prior but suppressed by another represent qualitatively different mass and spin characteristics, which highlights the failure of any of these intrinsic prior choices to be truly uninformative when inferring GW190521 source properties.}
    \label{fig:q_chieff_mtot_allpriors}
\end{figure}

\begin{figure}
    \centering
    \includegraphics[width=\linewidth]{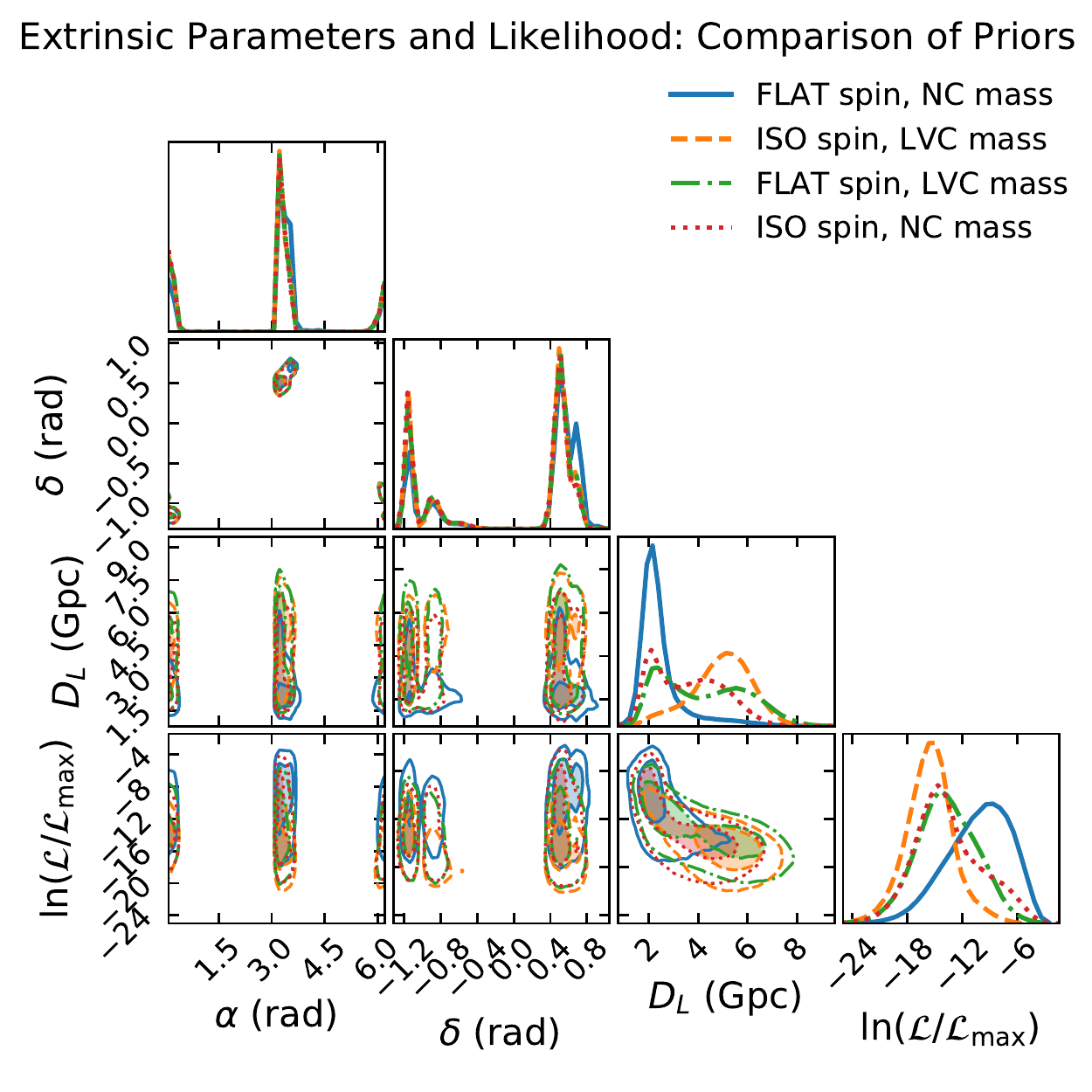}
    \caption{Sky location, luminosity distance, and likelihood compared across priors. The distributions of right ascension ($\alpha$) and declination ($\delta$) are robust to intrinsic prior choices, with the island at $(\alpha, \delta) \approx (3.3, 0.5)$ radians having a mild advantage in both posterior and likelihood. This consistency gives evidence that the different solutions are not just an artifact of unlucky antenna orientation. The priors do, however, lead to markedly different likelihood and distance distributions, with likelihood increasing as the distance decreases. This illustrates a direct competition between the likelihood and geometric prior volume, although the distance effect is not sufficient to explain the degree to which GW190521 posteriors diverge from the information contained in the likelihood despite attempts to make the mass and spin priors uninformative.}
    \label{fig:sky_DL_Lrat_allpriors}
\end{figure}

\section{GW190521 Posteriors}\label{sec:posteriors}

\subsection{Sampler Specifications}

We perform independent parameter estimation under each of the four sets of priors using the \textit{PyMultiNest} library \cite{pymultinest} powered by \textit{MultiNest} \cite{multinest_Feroz2013}, which is a nested sampling algorithm designed to be effective for multi-modal posteriors \cite{multinest_orig_Feroz2007, multinest_openFeroz2008}. The sampler computes the evidence integral using a user-specified number of ``live points" to draw equal-weight samples until a user-specified absolute error tolerance in the log evidence is achieved. The need for more costly settings than a typical GW event's PE was already reported by \citet{nitz2020_game_over} for GW190521, and we found this to be true both here and for the multi-modal posteriors of GW151226 \cite{boxingday2021}. We report results with 20000 live points ($\gtrsim$10 times more than required for a typical event) and a tolerance value of 0.05 (10 times more stringent than required for a typical event) across all priors.

We compute the likelihood using inspiral-merger-ringdown (IMR) waveforms generated by the IMRPhenomXPHM phenomenological approximant \cite{xphm_paper}. This includes the effects of higher-order multipole modes up to $\ell = 4$ and precessing spins over a broad range of mass ratios and spin magnitudes. We use the default precession implementation with all available modes (i.e., twisting up the co-precessing frame modes $(\ell, \abs{m}) \in \lbrace (2, 2), (2, 1), (3, 3), (3, 2), (4, 4) \rbrace$ with \textit{PrecVersion}=223). It is the same waveform model used by \citet{nitz2020_game_over} except for some recent updates to the precession implementation in the extreme mass ratio regime, and it is comparable to the phenomenological models used in the LVC analysis. Our whitened detector strain data is obtained by reprocessing the LVC data \cite{lvc_event_GW190521} and estimating the power spectral density (PSD) with drift correction using the methods of \citet{psd_drift}. Variance-weighted inner products are computed efficiently with relative binning \cite{relative_binning}.

\subsection{Posterior Sampling Results}

\begin{figure}
    \centering
    \includegraphics[width=\linewidth]{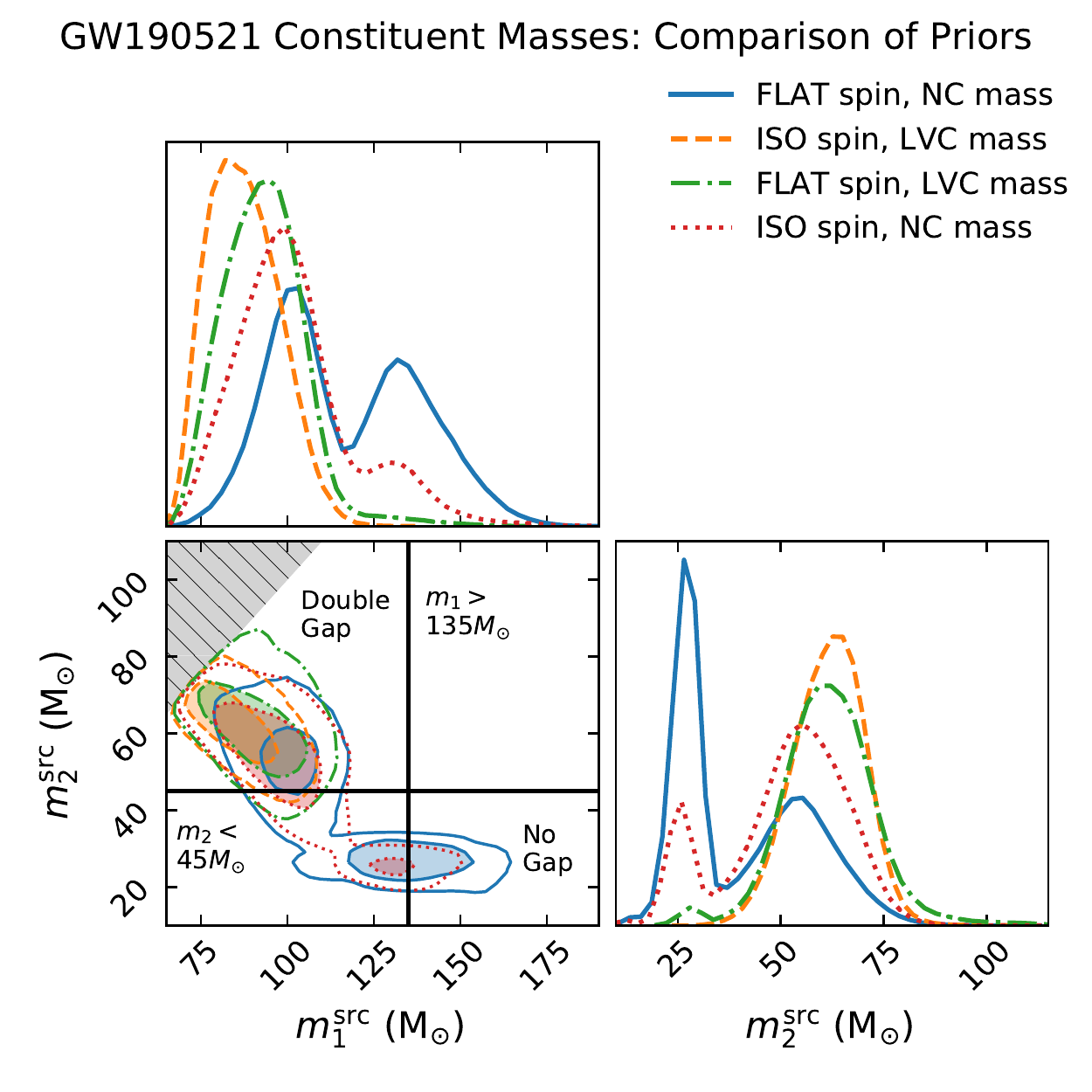}
    \caption{Constituent BH source-frame mass posteriors compared across priors, with a mass gap of [45, 135] $\msun$ indicated by black lines (the shaded region $m_2 > m_1$ is excluded by definition). In the 2-dimensional plots, the contours mark 50\% and 90\% confidence intervals. We see that the qualitatively different scenarios of mass ratio and spin highlighted by different priors also translate to different possible numbers of BH masses inside the gap. Posteriors under the LVC mass prior strongly prefer the double gap scenario, whereas the NC mass prior opens up the possibility of moving one or both BHs outside the mass gap.}
    \label{fig:m12_allpriors}
\end{figure}

\begin{figure}
    \centering
    \includegraphics[width=\linewidth]{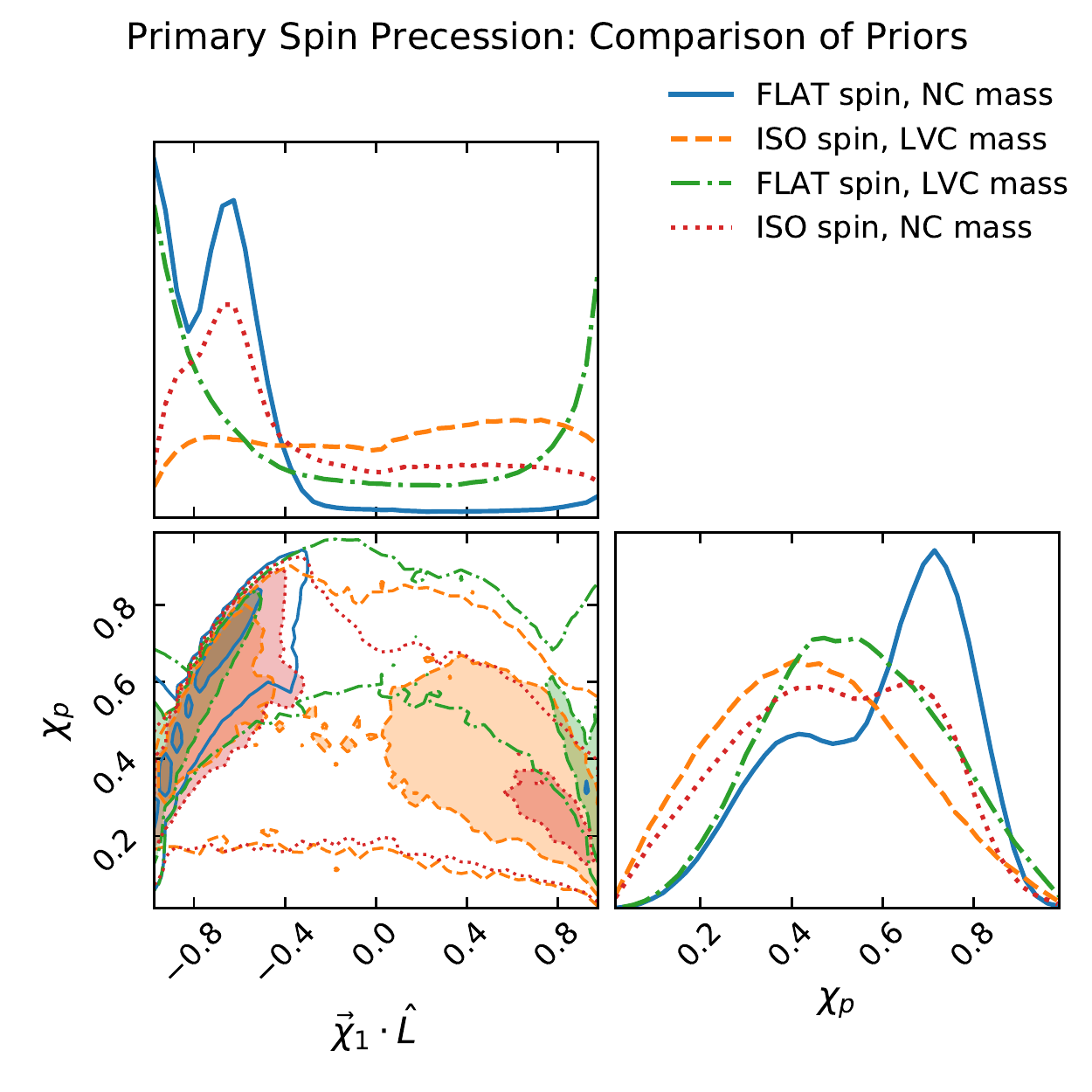}
    \caption{Precession of the primary BH spin examined through measurement of primary spin tilt cosine, $\vchi_1 \cdot \hat{L}$, and effective precession parameter, $\chi_p$. Focusing on each pair of posteriors with the same spin prior, we see the NC mass prior leads to increased preference for particular values of these precession parameters relative to the posteriors under the LVC mass prior in both cases. Although neither the spin tilt nor $\chi_p$ is very well-measured, this motivates us to investigate the possibility that the IMBH scenario (with support under the NC mass prior) requires a precessing primary in contrast to the lack of strong evidence for precession in the mass gap scenario.}
    \label{fig:s1inplane_priors}
\end{figure}

Posteriors from the ISO spin prior and LVC mass prior are largely consistent with the posteriors obtained by LVC using phenomenological approximants \cite{lvc_event_GW190521}, with most of the posterior weight contained in the region of primary mass less than twice the secondary mass and effective spin close to zero. Under the FLAT spin prior the peak near zero effective spin shifts up to $\chieff \sim 0.2$ and we find broad support for negative effective spin (Figure~\ref{fig:q_chieff_mtot_allpriors}). In Figure~\ref{fig:sky_DL_Lrat_allpriors} we see that this produces a second luminosity distance peak near 2 Gpc in addition to the peak found by LVC near 5 Gpc. The negative effective spin solution branch requires a closer distance to maintain comparable SNR to the non-negative branch due to the decrease in intrinsic luminosity at negative effective spin relative to positive spin. This effect is more pronounced for heavier systems, and the particular relevance to IMBH detection was recently explored by \citet{intrinsic_obs_dist_by_chieff_mtot_mehta2021observing}. \changed{There is a similar correlation between distance and mass ratio because, for fixed total mass and effective spin, equal-mass mergers have a luminosity advantage over extreme mass ratios.} In Figure~\ref{fig:m12_allpriors} we see the LVC mass priors favoring BHs inside the mass gap under both spin priors, although the FLAT spin prior has some support for a secondary mass below the gap. We do not find strong evidence for precession under the LVC mass prior, nor is there sufficient evidence to exclude the possibility (see Figure~\ref{fig:s1inplane_priors}).

Under the NC mass prior the story is very different. There is a large peak at a mass ratio of $q \sim 0.2$, as well as a smaller peak smeared across $q < 0.1$ that is enhanced by the FLAT spin prior, and regardless of spin prior there is a preference for negative effective spin (Figure~\ref{fig:q_chieff_mtot_allpriors}). We see that the solutions with higher likelihood are of the type uncovered by \citet{gap_stradler_fishbach2020} and \citet{nitz2020_game_over}, where one or both BHs is outside of the mass gap. There is also evidence of primary BH spin components in the plane of the orbit, which can be seen in the enhancement of a favored primary spin tilt when comparing the NC mass prior to the LVC mass prior for either of the spin priors, as shown in Figure~\ref{fig:s1inplane_priors}.

A large fraction of these unequal-mass, negative effective spin solutions have a primary spin at an extreme tilt with respect to the orbital angular momentum ($\sim$45 degrees from anti-alignment). We present this illustration of precession through spin tilt rather than the commonly used variable $\chi_p$ because the physical interpretation is clear. It should be noted however that neither $\chi_p$ nor the primary spin tilt is particularly well-measured here. An important step in the near future will be finding a consistent way to quantify precession, such as the work of \citet{generalized_chi_precession2021} to design new effective precession parameters for capturing the in-plane spin degrees of freedom over different timescales.

The dependence on intrinsic parameter priors makes it difficult to interpret the GW190521 source parameters without mass and spin priors that are assumed to describe the true astrophysical population from which the system originated. Bayesian inference is an excellent way to estimate parameters when one has a good understanding of the prior distribution, but without such knowledge the priors become a liability. However, without any assumptions about the astrophysical origin we can still use the fact that we have sampled a broad range of priors to construct a map of the GW190521 likelihood manifold.

\section{Likelihood Mapping}\label{sec:likelihood}

\subsection{Likelihood Maximization}\label{sec:maximization}

\begin{figure}
    \centering
    \includegraphics[width=\linewidth]{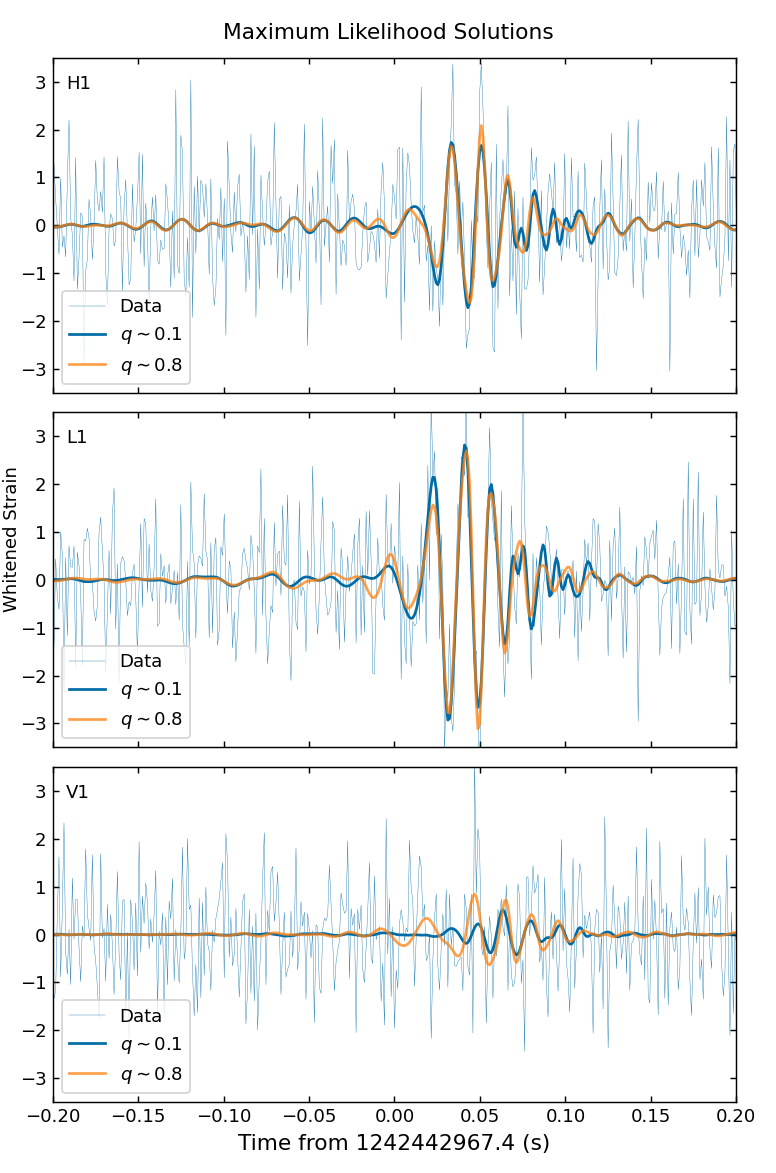}
    \caption{Maximum likelihood waveforms from numerical maximization over the unequal-mass and equal-mass regions. The equal-mass solution (orange) has source frame masses $(130, 110) \msun$ and an effective spin of 0.36, with primary spin magnitude 0.97 at a tilt of 86 degrees. The unequal-mass solution (blue) has source frame masses $(187, 16) \msun$ and an effective spin of $-0.35$, with primary spin magnitude 0.99 at a tilt of 107 degrees. Both lie in the upper tail of the total mass distribution and have a primary with nearly maximal spin at an extreme tilt with respect to the orbital angular momentum. The maximum likelihood ratio favors the unequal-mass solution over the equal-mass to a lesser degree than the average likelihood in each peak, with $\ln( \mathcal{L}_{\rm{max}}(q > 0.3) / \mathcal{L}_{\rm{max}}(q < 0.3) ) \approx -2$.}
    \label{fig:bestfit_wf}
\end{figure}

The simplest way to begin this mapping process is to maximize the likelihood over the entire parameter space. The maximum likelihood solution is the set of parameters giving the largest matched-filter SNR. On the one hand, since the data is noisy and some values of geometric parameters are less likely than others \textit{a priori}, the maximum likelihood solution can be misleading. On the other hand, sometimes the data speaks so loudly that it cannot be ignored. This was the case for the unequal-mass solution identified by \citet{nitz2020_game_over} under the previous version of the waveform model, which had a maximum likelihood $\sim$e$^{12}$ larger than the equal-mass peak. Using that same version we were able to reproduce the peak reported by \citet{nitz2020_game_over} for the unequal mass region, and the FLAT spin prior led to improvement in the equal mass peak's likelihood relative to the \nitz samples, giving a likelihood ratio of $\sim$e$^{10}$ between the two maxima.

However, in agreement with other recent GW190521 studies \cite{quasinormal_ringdown_PEcorrection_nitzCapano2021, new_scoop_waveform_model_comparison_xphm2021} we no longer find such a dramatic difference under the latest waveform model version. To search for a similarly exceptional solution with the updated waveform model, we maximized the likelihood over different grids throughout parameter space, refining to smaller sub-intervals of $M \in [100, 360] \msun, \, q \in [0.04, 1], \, \chieff \in [-0.99, 0.99]$ as we found better local maxima. We also ran the posterior sampler over restricted regions in addition to the full PE, but the result from the previous waveform model version was not recovered. With the updated IMRPhenomXPHM we found a reduction of the maximum log likelihood by $\sim$5 in the unequal-mass peak versus an increase of $\sim$3 in the equal-mass peak relative to the former version. This reduces the maximum log likelihood difference to $\sim$2 between the peaks. Considering uncertainties in waveform modelling and PSD estimation, and the irreducible noise which randomly boosts some solutions relative to others, this new peak value is less compelling than the likelihood advantage under the previous version. The whitened waveforms from the maximum likelihood parameters for $q < 0.3$ (unequal-mass) and $q > 0.3$ (equal-mass) regions of parameter space are shown in Figure~\ref{fig:bestfit_wf}.

Although the likelihood's maxima are comparable between the regions, the maximum likelihood solution for the equal-mass region has more fine-tuned support. In particular, the unequal-mass peak is hard to find in sampling but easier in maximization because it has dense support near the maximum, whereas our maximization over sub-intervals in the equal-mass region found only a few narrow upward fluctuations producing likelihood comparable to the maximum. The same structure appears in the distributions of posteriors samples (seen most clearly in the mass ratio vs. likelihood panel of Figure~\ref{fig:qchi_Lrat_DL_by_ngap_mode}), where the sampler identified dense clusters of high-likelihood solutions near the peak of the unequal-mass region, whereas the equal-mass region is populated primarily by lower-likelihood solutions with sporadic points of increased likelihood. 

\changed{This average likelihood difference between different regions of parameter space across all sampling priors is shown in Figure~\ref{fig:m12_lnL_DL_by_qchi_mode}, where we combine all the posterior samples into an ensemble for mapping likelihood structures. When choosing solutions representative of different regions of parameter space, the local maxima of the likelihood have the attractive feature of being prior-independent. However, the maximum likelihood does not apply any penalty for fine-tuning: due to parameter degeneracies, different solutions (in the sense of waveform shapes) have different amounts of phase space volume associated.} 

\changed{The quantification of this volume is prior dependent, and hence arbitrary to some degree, but altogether absent from the maximum likelihood criterion. We also note that at least the prior on extrinsic parameters is well justified. Figure~\ref{fig:m12_lnL_DL_by_qchi_mode} shows that this effect is quite important for the case of GW190521: although the likelihood's maxima are comparable between the regions, the average likelihood in each are quite different. Moreover, the likelihood has a sizeable anticorrelation with the source's distance, with better fitting solutions being more finely tuned to a smaller distance. Motivated by these observations, in the following section we devise another criterion to identify representative solutions that includes information from the phase space volume.}

\subsection{Characteristic Solutions by Mass Ratio and Effective Spin}\label{sec:q_chi_modes}

\begin{figure*}
    \centering
    \subfloat[]{
    \includegraphics[width=.5\linewidth]{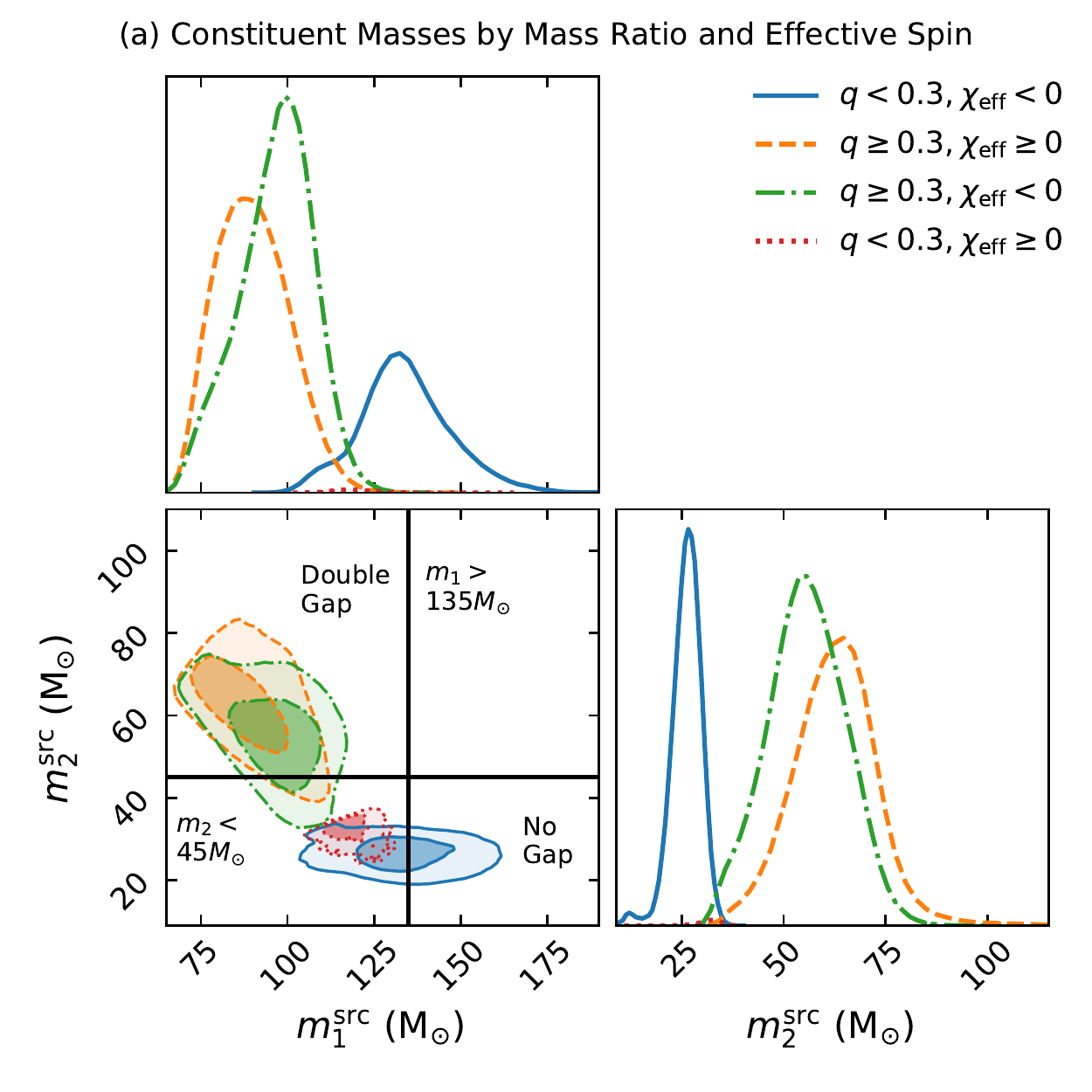}}
    \subfloat[]{
    \includegraphics[width=.5\linewidth]{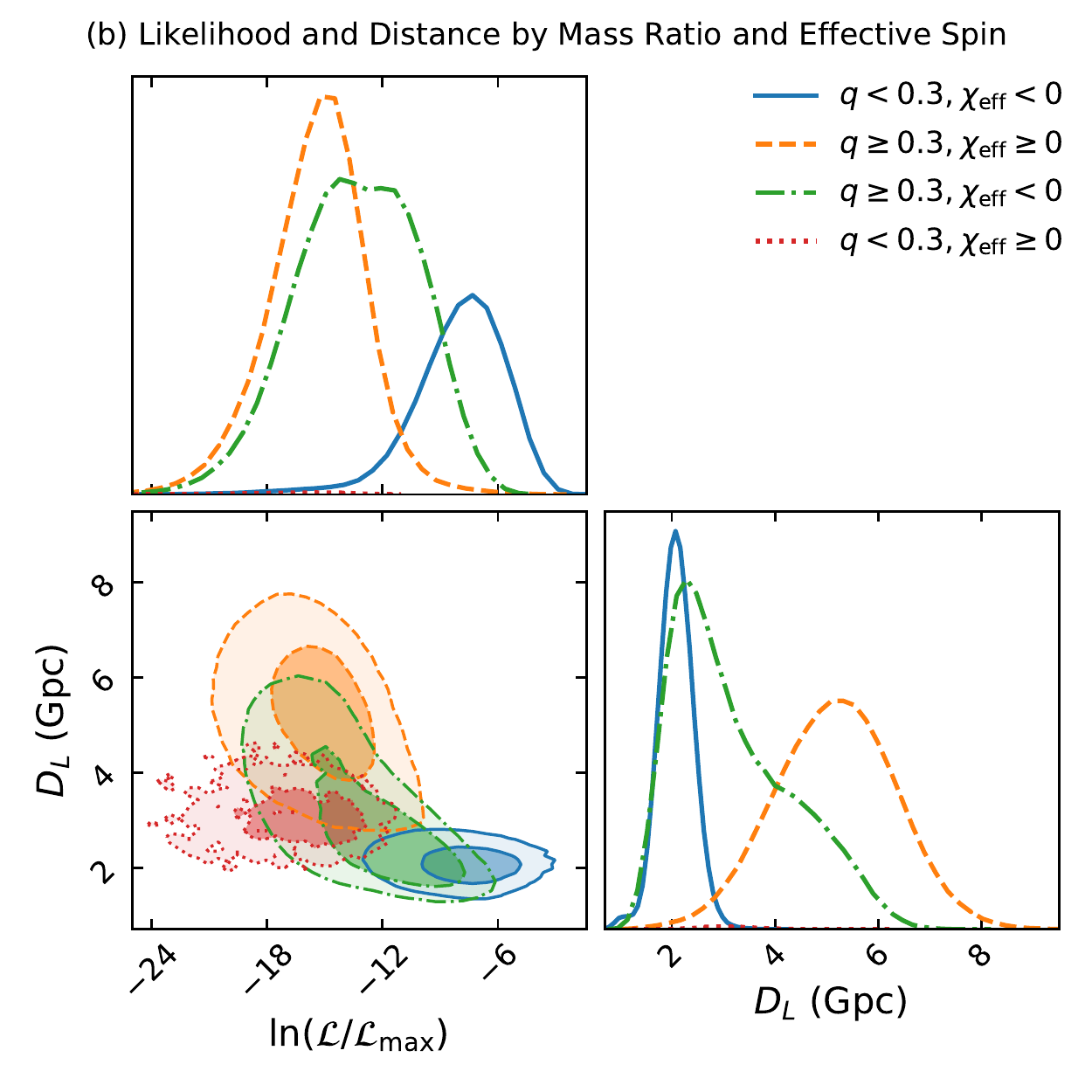}}
    \caption{Constituent BH masses (a), distance and likelihood (b) compared between quadrants of mass ratio and effective spin \changed{for the ensemble of all samples combined across the four sampling priors}. The mass gap of [45, 135] $\msun$ is indicated by black lines. We see at least two peaks associated to different characteristic distances and likelihoods, and the solutions with comparable masses but negative effective spin (green) appear to trace out waveform degeneracy between the peaks. The 1-D plots share the same normalization across ensembles, showing the negligible contribution of the unequal-mass region with non-negative spin. In the 2-D plots, the contours mark the 50th and 90th percentiles. Through this breakdown of solutions we construct a map of the likelihood structures suggested by Figures~\ref{fig:q_chieff_mtot_allpriors}, \ref{fig:m12_allpriors}, and \ref{fig:sky_DL_Lrat_allpriors}, but which we were unable to fully examine through any single choice of prior. Both the unequal-mass solutions with negative effective spin (blue) and equal-mass solutions with non-negative effective spin (orange) have fairly well-specified characteristic distances and likelihood, which give the latter a clear advantage in prior volume and the former an even larger advantage in fitting the data. The mass gap can only be avoided where $q \leq 0.3, \chieff < 0$ (which places the secondary mass at $\sim$25 $\msun$ independent of the primary mass, guaranteeing at least one BH outside the gap), whereas $q > 0.3, \chieff \geq 0$ solutions lie almost exclusively in the double gap region.}
    \label{fig:m12_lnL_DL_by_qchi_mode}
\end{figure*}

If we combine the samples from all our priors, we must abandon the inference question (``What are the source parameters of GW190521?'') and instead adopt the less powerful but more answerable question: ``Considering all the regions of parameter space where our physical model can consistently produce a good fit to the data, how does fixing some subset of the parameters constrain the other parameters?". This helps us understand how the various approximate physical degeneracies between waveforms with different parameters are navigated in this particular realization of noise, under the constraint that we match a putative signal which may fall into one of a few (possibly overlapping) regions of parameter space. In Figure~\ref{fig:m12_lnL_DL_by_qchi_mode} we can see that the only solution with support for the scenario that both BHs have masses outside the mass gap is the unequal-mass solution with negative effective spin.

We also see the likelihood peaks selecting for different luminosity distances, with the negatively spinning solutions preferring less distant sources. We expect this to some degree from the interplay between the shape of the PSD and the lower intrinsic loudness of negative effective spin mergers relative to zero and positive effective spin \cite{orbital_hangup_Campanelli2006}, especially for high-mass mergers \cite{intrinsic_obs_dist_by_chieff_mtot_mehta2021observing}. All regions of parameter space have a remarkably consistent distribution of sky positions in this breakdown, which helps assure us that these different solution regions are not simply different polarizations and antenna projections of the same set of parameters. This is a useful breakdown of the parameter space because, in addition to the evident separation of solutions by mass ratio and effective spin, there has been much investigation into the presence or lack of mergers with negative effective spin, and how this relates to selection effects and population modeling \cite{lvc_POP-2, LVC_pop1_o2, ias_popO2_Roulet_2020, newpop_o3a_Roulet2021, lowestSNR_O2_ias_mit2020ysn, field_binary_no_chieff_pop2020, impact_of_priors_Vitale2017, BBHpop_MchiCorr_Roulet_2019, mass_chieff_trends_pop_farr2020mlb}.

In Table~\ref{table:soln_mode_examples} we consider representative samples for the following intrinsic parameter space regions:
\begin{equation}\label{eqn:p_defs_qchi}
\begin{split}
    \pint^{U-} &\equiv \lbrace q \leq 0.3, \chieff < 0 \, \big| \, \mathcal{L} / \mathcal{L}_{\rm max} > \rm{e}^{-18} \rbrace \\
    \pint^{E-} &\equiv \lbrace q > 0.3, \chieff < 0 \, \big| \, \mathcal{L} / \mathcal{L}_{\rm max} > \rm{e}^{-18} \rbrace \\
    \pint^{E+} &\equiv \lbrace q > 0.3, \chieff \geq 0 \, \big| \, \mathcal{L} / \mathcal{L}_{\rm max} > \rm{e}^{-18} \rbrace
\end{split}
\end{equation}
For GW190521, these intrinsic parameter space regions can be associated to characteristic likelihoods and distances $(\bar{D}, \bar{\mathcal{L}})$. The likelihood ratio cutoff of $\sim10^{-8}$ relative to the maximum is to prevent our notion of characteristic distance and likelihood from being influenced by solutions that do not fit the data well.  We do not discuss further the region of unequal-mass solutions with positive effective spin because we did not find a comparable likelihood peak in that quadrant. In each of the parameter space regions above, the data identify a cluster of high-likelihood solutions in the vicinity of some characteristic set of parameters.

\begin{figure}
    \centering
    \includegraphics[width=\linewidth]{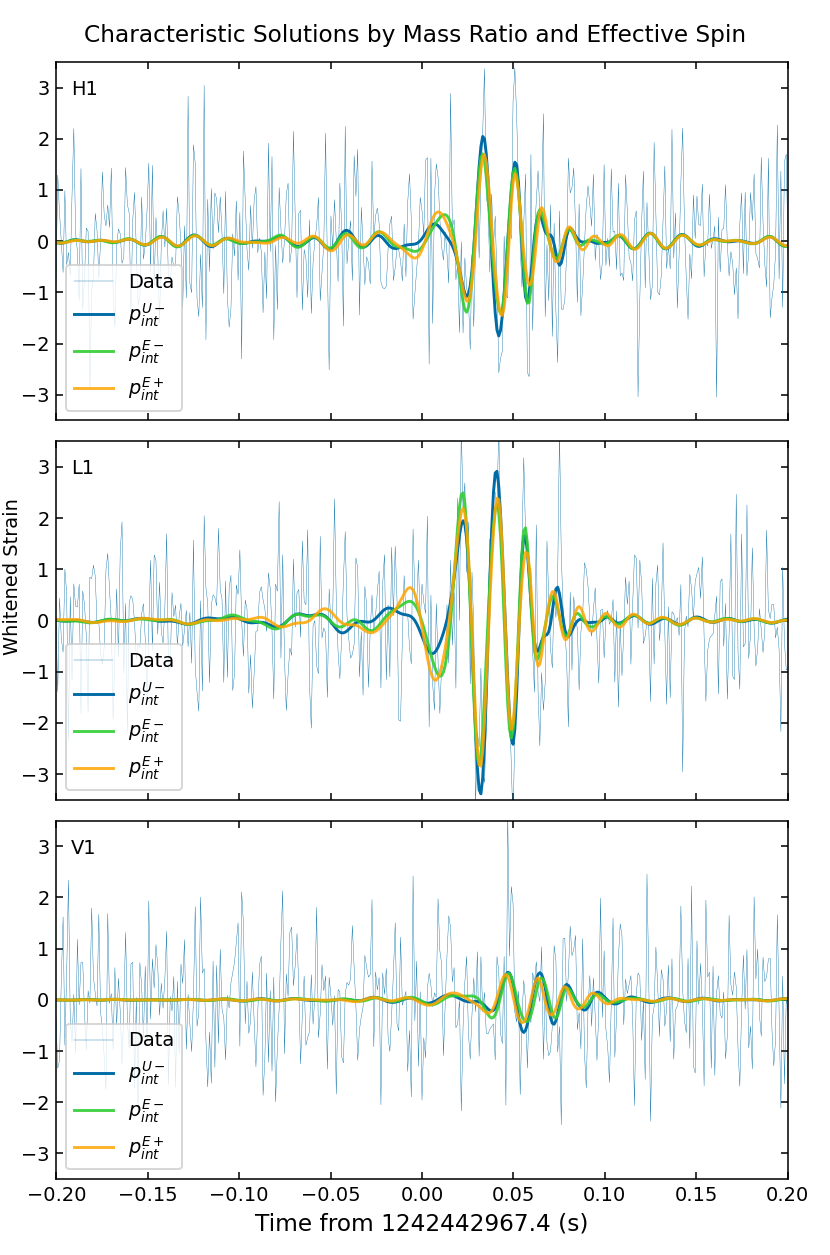}
    \caption{Whitened strain at each detector for the representative samples given in Table \ref{table:soln_mode_examples}. \changed{Note that these parameters are illustrative and cannot be interpreted with the same statistical rigor as, e.g., maximum likelihood solutions.}}
    \label{fig:char_wf}
\end{figure}

The representative samples in Table~\ref{table:soln_mode_examples} are a compromise between the noisy likelihood peaks and the prior-driven posteriors. To balance these effects, we consider samples within a standard deviation from the medians of the parameter regions $\pint^{U-}, \pint^{E-}, \pint^{E+}$ and restrict to the patch of sky that is favored across all priors: right ascension near 3.3 radians and declination near 0.5 radians (see Figure~\ref{fig:sky_DL_Lrat_allpriors}). For each region we select 100 samples at random and compute the matches between each pair of waveforms. The representative is the sample with the largest sum of squared matches over the set. \changed{This maximization mitigates the randomness associated to the stochasticity of the sampling and, to some degree, the influence of our specific choice of priors.} The whitened waveforms and data are plotted at each detector in Figure~\ref{fig:char_wf}. Keep in mind that these are just representative samples, so the parameters in Table~\ref{table:soln_mode_examples} can only be used for qualitative illustration and order-of-magnitude estimation.

\begin{table}
\begin{center}
\setlength\tabcolsep{3pt}
\begin{tabular}{lcccc}
\specialrule{.1em}{.05em}{.05em} 
\specialrule{.1em}{.05em}{.5em} 
Samples at  $\, (\alpha, \delta) \approx (3.3, 0.5) \, \, \, \,$  & $\pint^{U-}$ & $\pint^{E-}$ & $\pint^{E+}$ \\ 
\specialrule{.1em}{.3em}{.5em}
$\mathcal{M}^{\rm det}$ [$M_{\odot}$] & $66$ & $96$ & $122$ \\[4pt]
$q   $  & $0.18$  & $0.54$ & $0.90$ \\[4pt]
$m_1$ [$M_{\odot}$]      & $139$ & $105$ & $85$ \\[4pt]
$m_2$  [$M_{\odot}$]   & $25$ & $56$ & $77$ \\[4pt]
$\chieff$ & $-0.47$ & $-0.24$ & $0.18$ \\[4pt]
$\vchi_1 \cdot \hat{L}$ & $-0.65$ & $-0.84$ & $0.62$ \\[4pt]
$| \vec{\chi}_1 |$ & $0.93$ & $0.77$ & $0.43$ \\[4pt]
$D_L$ [Gpc] & $2.1$ & $2.5$ & $4.7$ \\[4pt]
$VT$ ratio & $0.2$ &: $\, \, 0.6$ &: $\, \, \, 1$ \\[4pt]
$D_{\rm{com}}^3 \mathcal{L}$ ratio  & $1$ & : $0.03$ &: $0.01$ \\[4pt]
\specialrule{.1em}{.05em}{.05em} 
\specialrule{.1em}{.05em}{.05em} 
\end{tabular}
\setlength{\belowcaptionskip}{-15pt}
\caption{Source parameters representing characteristic solutions in each of the likelihood peaks (with parameter space regions defined in Equation~\eqref{eqn:p_defs_qchi}), restricting to the patch of sky that is favored across all priors and solution regions. The chirp mass is detector-frame and the constituent masses are source-frame. The $VT$ of each solution is integrated numerically following Equation~\eqref{eqn:vt}, where intrinsic luminosity distance, $D_{L, \rm{max}}$, is computed by numerical maximization of $\sqrt{\overlap{h_+}{h_+}}$ over inclination and orbital phase using a PSD constructed to represent high-sensitivity single detector operation during O3a (averaging procedure described in \cite{ias_template_bank_PSD_roulet2019}), and defined for an SNR threshold of 10 in this fiducial detector. We also compute the ratio of $D_{\rm{com}}^3 \mathcal{L}$ to give some sense of the competition between the likelihood and the geometric prior volume, with the likelihood advantage of $\pint^{U-}$ overpowering the intrinsic luminosity advantage of $\pint^{E+}$ before considering the mass and spin priors.}
\label{table:soln_mode_examples}
\end{center}
\end{table}

Given these representative samples, we compare the characteristic sensitive $VT$ of each solution region. If we believe the universe produces equal numbers of each type of merger per unit comoving volume-time, then the ratio of expected number of detections is the ratio of the maximum comoving volume-time for which a system with those intrinsic parameters is expected to exceed some SNR threshold. The threshold drops out of the ratio but is necessary for getting absolute distances and their cosmological redshifts, so we use an SNR threshold of 10. Thus we compare
\begin{equation}\label{eqn:vt}
    VT(\pint) \, = \int\limits_0^{D_{L, \rm{max}}(\pint)} \frac{4 \pi D_L^2}{(1 + z)^4} \bigg( 1 - \frac{D_L}{1 + z} \frac{\rmd z}{\rmd D_L} \bigg) \rmd D_L ,
\end{equation}
where again the redshift computations assume a $\rm{\Lambda C D M}$ cosmology with Planck15 results \cite{cosmology_planck2015}. We might worry that this particular data calls for tuning extrinsic parameters like inclination such that the characteristic distance for GW190521 is much less than the maximum distance at which a system with those intrinsic parameters is observable (this effect is discussed in, e.g., \cite{intrinsic_obs_dist_by_chieff_mtot_mehta2021observing}). To account for this we define the sensitive $VT$ in terms of intrinsic luminosity distance, $D_{L, \rm{max}}(\pint)$, which is proportional to $\sqrt{\overlap{h_+}{h_+}}$ maximized over all extrinsic parameters for a given set of intrinsic parameters.

Using a fiducial O3a PSD (constructed as in \cite{ias_template_bank_PSD_roulet2019} to give a representation of high-sensitivity single detector operation) with an SNR threshold of 10, we estimate $D_{L, \rm{max}}$ to be $\sim$3.2 Gpc for $\pint^{U-}$, $\sim$5.8 Gpc for $\pint^{E-}$, and $\sim$8.7 Gpc for $\pint^{E+}$. This gives a rough estimate for the comoving volume-time ratio, $VT$ of $\pint^{U-} \,:\, \pint^{E-} \,:\, \pint^{E+} \,\, \sim \,\, 0.2 \,:\, 0.6 \,:\, 1$. This is only a zeroth-order approximation, so it should not be over-interpreted. At most we can make very mild statements by assuming a BBH population with constant merger rate densities $R_{U-}, R_{E-}, R_{E+}$ for the characteristic systems. For instance, if $R_{U-} / R_{E+} \ll 10$ then it would be surprising to have observed a system in $\pint^{U-}$ before observing a system in $\pint^{E+}$. Since GW190521 would mark the first observation of its kind no matter which of the possible solutions we choose to believe, this type of statement helps to categorize our degree of surprise at each conclusion given some astrophysical population model.

\changed{We see that, for the representative samples in Table \ref{table:soln_mode_examples}, the characteristic likelihood advantage of $p_{U-}$ over $p_{E+}$ is opposed by a phase space volume difference coming from the intrinsic luminosity advantage of $p_{E+}$ over $p_{U-}$. To get a rough idea of how the effects compare locally, consider the ratio of single-sample contributions to the evidence before specifying rate densities (i.e., in absence of astrophysical mass and spin priors): the likelihood ratio multiplied by the geometric prior ratio from the extrinsic parameters. The only extrinsic parameter whose phase space volume appreciably differs between $p_{U-}$ and $p_{E+}$ is the distance. Thus we include the ratio of $D^3_{com} \mathcal{L}$ between the samples in Tables \ref{table:soln_mode_examples} and \ref{table:gap_mode_examples}. This is a proxy for the relative local evidence contributions of samples before mass and spin priors are considered, providing a rough scale comparison for the effects of characteristic likelihood and distance differences that we see in the likelihood maps. The representative samples in Table~\ref{table:soln_mode_examples} give $D^3_{com} \mathcal{L}$ ratios of $\pint^{U-} \,:\, \pint^{E-} \,:\, \pint^{E+} \,\, \sim \,\, 1 \,:\, 0.03 \,:\, 0.01$. At these characteristic points we find that, compared to the structure of the likelihood manifold, the extrinsic phase space volume is a subdominant effect.}

\subsection{Characteristic Solutions by Mass Gap}\label{sec:ngap_modes}

\begin{figure}
    \centering
    \includegraphics[width=\linewidth]{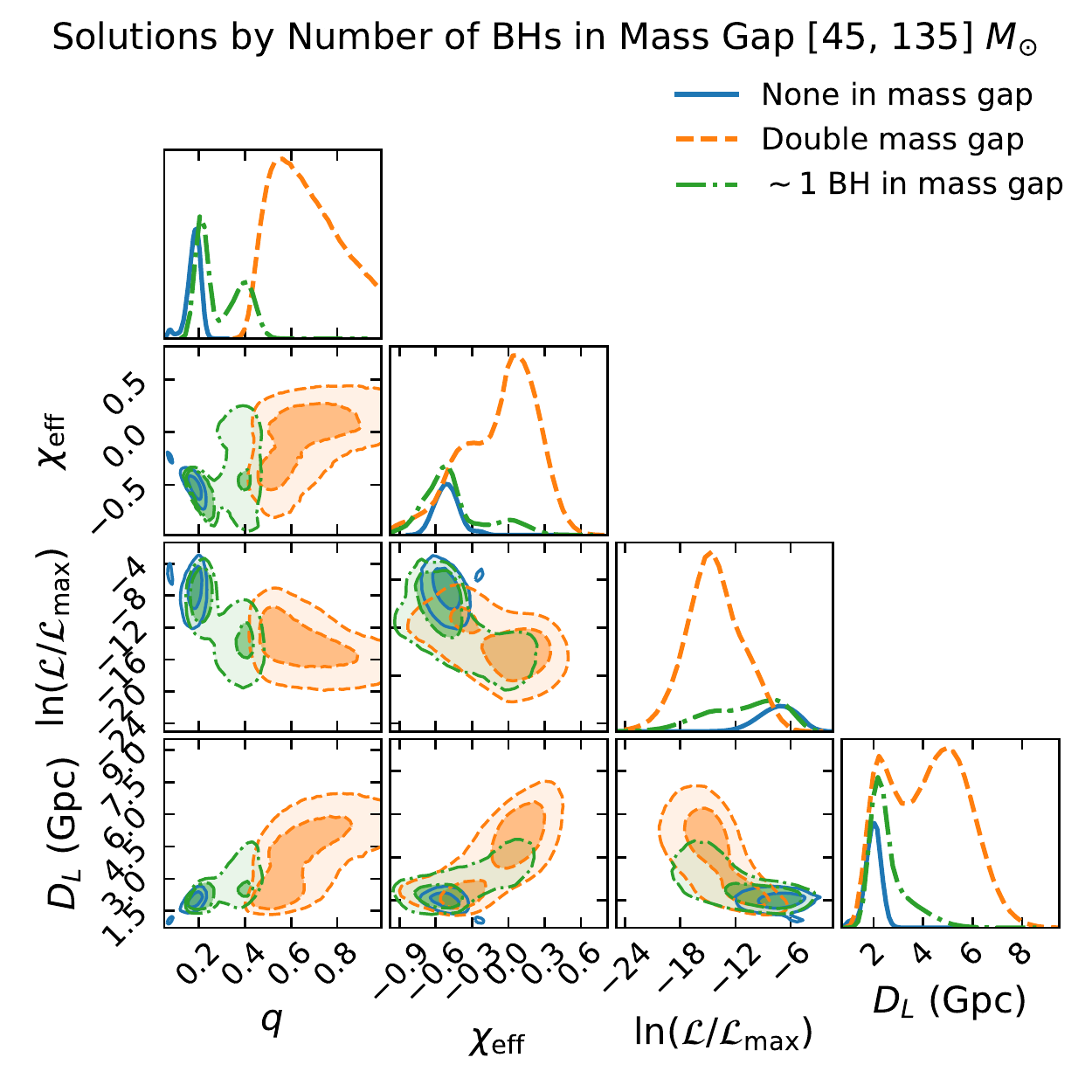}
    \caption{Mass ratio, effective spin, likelihood and distance compared between regions with 0, 1 or 2 BH masses falling in the gap of [45, 135] $\msun$ \changed{for the ensemble of all samples combined across the four sampling priors}. In the 2-D plots, the contours mark 50th and 90th percentiles. The 1-D plots share the same normalization across ensembles. We see that there are many more samples making up the scenarios with BHs in the mass gap, but these distributions are fattened by low-likelihood solutions that entered the ensemble through priors which gave a pathological representation of the maximum likelihood, and therefore also of the minimum likelihood required to be relevant. The likelihood advantage of the no gap scenario over the double gap scenario is, however, less pronounced than that of $\pint^{U-}$ over $\pint^{E+}$. This is because the double gap samples also include a contribution from the negative effective spin branch of the equal-mass region, where we find the higher-likelihood solutions at closer distances. }
    \label{fig:qchi_Lrat_DL_by_ngap_mode}
\end{figure}

 In a similar qualitative sense we may want to know what spin and extrinsic characteristics are implied if we require that some number of BHs fall in a mass gap of $\sim$[45, 135] $\msun$. In Figure~\ref{fig:qchi_Lrat_DL_by_ngap_mode} we see that the scenario with both BHs falling outside this mass gap roughly maps onto the region $\pint^{U-}$, and so requires an unequal-mass system with a tilted primary spinning against the orbit at a luminosity distance $\sim$2 Gpc. The double mass gap scenario has two different peaks roughly separated between $\chieff$ being moderate and negative versus possibly positive but consistent with zero. These two solutions correspond to luminosity distances of $\sim$2.5 Gpc and $\sim$5 Gpc, respectively. We can again predict the observed distance effect through the relationship between intrinsic loudness and effective spin.

\begin{figure}
    \centering
    \includegraphics[width=\linewidth]{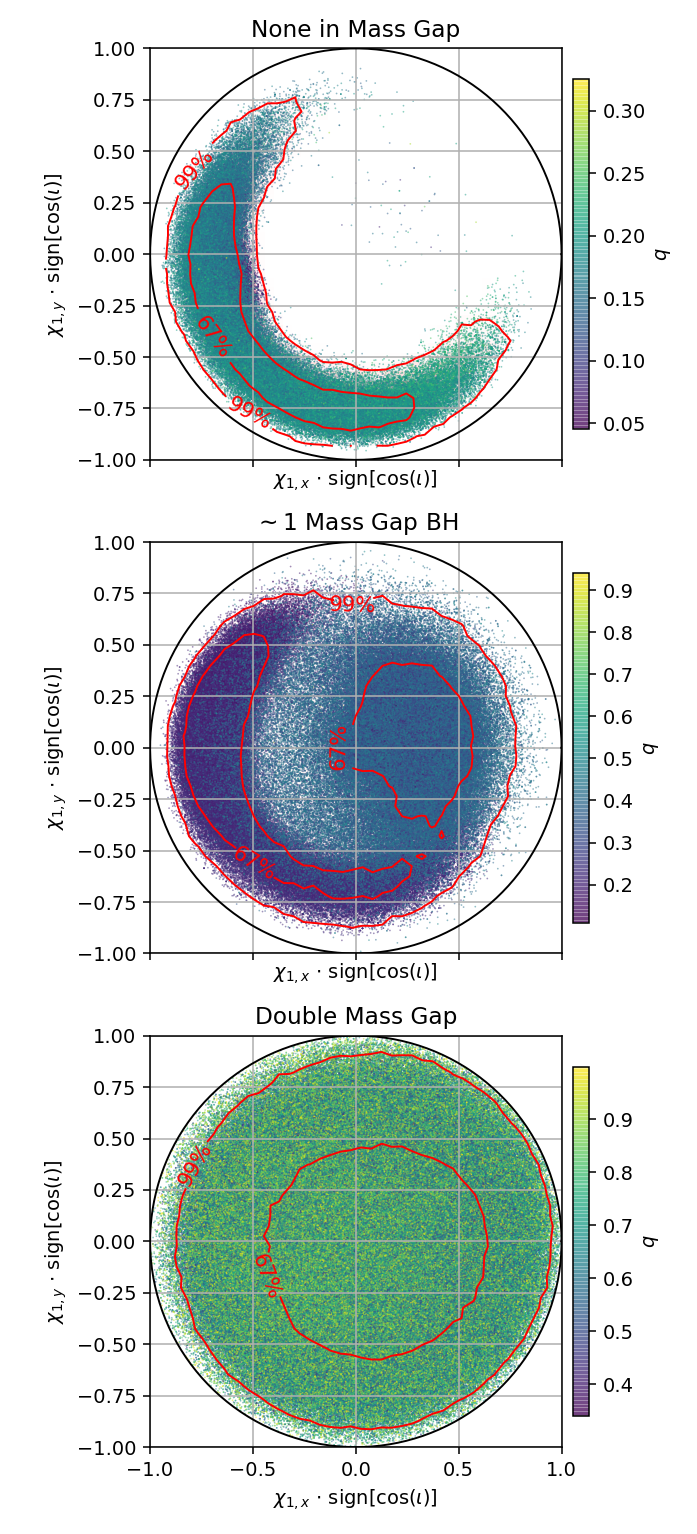}
    \caption{Primary BH in-plane spin over all samples for scenarios with 0, 1, or 2 BH masses inside the gap of [45, 135] $\msun$. Nonzero in-plane spin is associated to the case of no mass gap BHs at high significance, whereas the case for precession is indeterminate for the double gap scenario. Solutions with one BH mass in the gap make up a region of interpolation between the other two distributions. }
    \label{fig:s1inplane_gap}
\end{figure}

Somewhat less expected is the strong preference for a primary BH with in-plane spin components in the unequal-mass solution. In the top panel of Figure~\ref{fig:s1inplane_gap} we see that a primary spin with vanishing in-plane components is excluded at high confidence from the solution region with both BHs outside the gap. The bottom panel shows that the double mass gap scenario, on the other hand, is consistent with essentially any value of the in-plane spin components, making it hard to argue for or against a precessing primary. The solution with one BH in the gap (consisting almost entirely of scenarios with the smaller BH below the gap and the primary more than twice as massive) splits between a region of mass ratio near the unequal-mass peak of $q \sim 0.2$, which has an IMBH just below the upper edge of the gap, and a second region of $q \sim 0.5$ and primary mass $\lesssim 100 \msun$ that is qualitatively similar to the negative spin branch of the double mass gap solution.

\begin{figure}
    \centering
    \includegraphics[width=\linewidth]{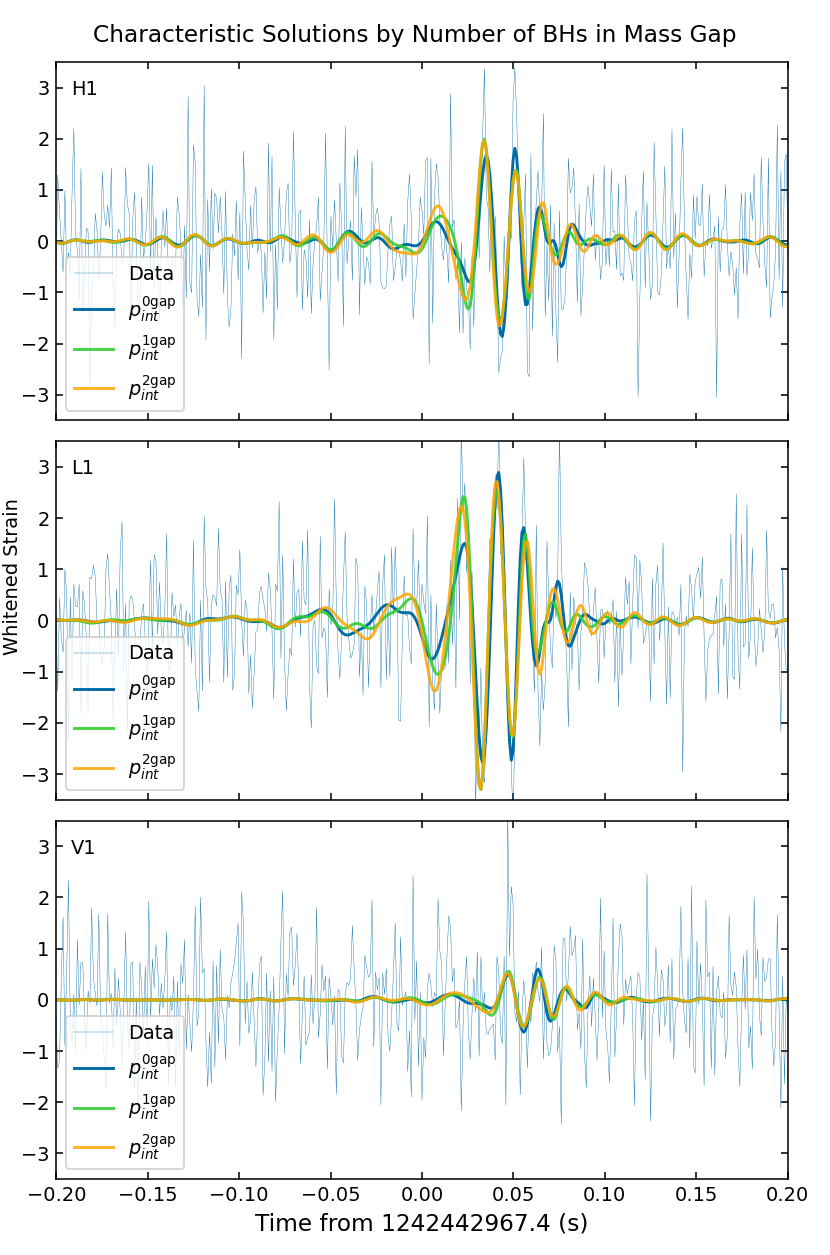}
    \caption{Whitened strain at each detector for the representative samples given in Table \ref{table:gap_mode_examples}. \changed{Note that these parameters are illustrative and cannot be interpreted with the same statistical rigor as, e.g., maximum likelihood solutions.}}
    \label{fig:gap_wf}
\end{figure}

From the likelihood mapping perspective, the ensemble with one BH in the gap is more like an interpolation from the lower-likelihood region of equal-mass solutions to the higher-likelihood region of unequal-mass solutions. The first column of panels in Figure~\ref{fig:qchi_Lrat_DL_by_ngap_mode} illustrates the role of this intermediary scenario as a bridge between two solution regions, and the high degree of uncertainty in placing the edges of the mass gap makes it reasonable to interpret it that way rather than imposing physical significance on the blurry boundary between certainly violating the gap and probably avoiding it.

\begin{table}
\begin{center}
\setlength\tabcolsep{3pt}
\begin{tabular}{lcccc}
\specialrule{.1em}{.05em}{.05em} 
\specialrule{.1em}{.05em}{.5em} 
Samples at  $\, (\alpha, \delta) \approx (3.3, 0.5) \, \, \, \,$  & $\pint^{0\rm{gap}}$ & $\pint^{1\rm{gap}}$ & $\pint^{2\rm{gap}}$ \\ 
\specialrule{.1em}{.3em}{.5em}
$\mathcal{M}^{\rm det}$ [$M_{\odot}$] & $70$ & $83$ & $126$ \\[4pt]
$q   $  & $0.19$ & $0.41$ & $0.81$ \\[4pt]
$m_1$ [$M_{\odot}$]      & $144$ & $100$ & $88$ \\[4pt]
$m_2$  [$M_{\odot}$]   & $27$ & $41$ & $71$ \\[4pt]
$\chieff$   & $-0.48$ & $-0.46$ & $0.26$ \\[4pt]
$\vchi_1 \cdot \hat{L}$ & $-0.64$ & $-0.76$ & $0.89$ \\[4pt]
$| \vec{\chi}_1 |$   & $0.92$ & $0.97$ & $0.53$ \\[4pt]
$D_L$ [Gpc]   & $2.1$ & $3.1$ & $5.4$ \\[4pt]
$VT$ ratio   & $0.2$ &: $\, \, 0.4$ &: $\, \, \, 1$ \\[4pt]
$D_{\rm{com}}^3 \mathcal{L}$ ratio   & $1$ &: $\, \, 0.4$ &: $0.02$ \\[4pt]
\specialrule{.1em}{.05em}{.05em} 
\specialrule{.1em}{.05em}{.05em} 
\end{tabular}
\setlength{\belowcaptionskip}{-15pt}
\caption{Source parameters representing solutions with 0, 1 or 2 BHs in the mass gap of [45, 135] $\msun$ (see Equation~\eqref{eqn:p_defs_ngap}), restricting to the patch of sky that is favored across all priors and solution regions. The chirp mass is detector-frame and the constituent masses are source-frame. The $VT$ of each solution is integrated numerically following Equation~\eqref{eqn:vt}. Intrinsic luminosity distance, $D_{L, \rm{max}}$, is computed by numerical maximization of $\sqrt{\overlap{h_+}{h_+}}$ over inclination and orbital phase using a PSD constructed to represent high-sensitivity single detector operation during O3a (averaging procedure described in \cite{ias_template_bank_PSD_roulet2019}), and defined for an SNR threshold of 10 in this fiducial detector. We also compute the ratio of $D_{\rm{com}}^3 \mathcal{L}$ to give some sense of the competition between the likelihood and the geometric prior volume, with the likelihood advantage of $\pint^{0\rm{gap}}$ overpowering the intrinsic luminosity advantage of $\pint^{\rm{2gap}}$ before considering the mass and spin priors. }
\label{table:gap_mode_examples}
\end{center}
\end{table}

Turning again to representative samples to get some rough intuition, we identify the following intrinsic parameter space regions (with mass values in units of $\msun$):
\begin{equation}\label{eqn:p_defs_ngap}
\begin{split}
    \pint^{0\rm{gap}} &\equiv \lbrace m_1 \geq 135 , m_2 \leq 45 \, \big| \, \mathcal{L} / \mathcal{L}_{\rm max} > \rm{e}^{-18} \rbrace \\
    \pint^{1\rm{gap}} &\equiv \lbrace m_1 < 135 , m_2 \leq 45 \, \big| \, \mathcal{L} / \mathcal{L}_{\rm max} > \rm{e}^{-18} \rbrace \\
    \pint^{2\rm{gap}} &\equiv \lbrace m_1 < 135 , m_2 > 45 \, \big| \, \mathcal{L} / \mathcal{L}_{\rm max} > \rm{e}^{-18} \rbrace
\end{split}
\end{equation}
We make the same cut in likelihood ratio to prevent the representative sample draws from being influenced by waveforms that do not fit the data well, and we disregard the region $m_1 > 135 \msun, m_2 > 45$ because it contains a negligible number of samples and no likelihood peak. Following the same random procedure for obtaining the representative samples in Table~\ref{table:soln_mode_examples}, we select representative samples from these mass gap regions. The parameters are given in Table~\ref{table:gap_mode_examples} and their whitened waveforms are plotted at each detector in Figure~\ref{fig:gap_wf}.

By the same methods as in Section~\ref{sec:q_chi_modes}, we estimate $D_{L, \rm{max}}$ to be 
$\sim$3.3 Gpc for $\pint^{0\rm{gap}}$, $\sim$4.8 Gpc for $\pint^{1\rm{gap}}$, and $\sim$9.0 Gpc for $\pint^{2\rm{gap}}$, which leads to an approximate comoving volume-time ratio, $VT$ of $\pint^{0\rm{gap}} \,:\, \pint^{1\rm{gap}} \,:\, \pint^{2\rm{gap}} \,\, \sim \,\, 0.2 \,:\, 0.4 \,:\, 1$. If we imagine a BBH population with constant merger rate densities $R_{0\rm{gap}}, R_{1\rm{gap}}, R_{2\rm{gap}}$ for the characteristic systems, then we would be surprised to have observed the gap-avoiding scenario before the double gap scenario if $R_{0\rm{gap}} / R_{2\rm{gap}} \ll 10$. \changed{On the other hand, using the same proxy for relative local evidence contributions in absence of rate densities as in the previous section, we find $D^3_{com} \mathcal{L}$ ratios of $\pint^{0\rm{gap}} \,:\, \pint^{1\rm{gap}} \,:\, \pint^{2\rm{gap}} \,\, \sim \,\, 1 \,:\, 0.4 \,:\, 0.02$. In this rough scale comparison between the effects of characteristic likelihood and distance differences evident in the likelihood maps, similar to $p_{U-}$ versus $p_{E+}$, we see that the likelihood advantage of the gap-avoiding scenario wins out over the intrinsic luminosity advantage of the double mass gap scenario for the representative samples in Table~\ref{table:gap_mode_examples}}.

\section{Discussion}\label{sec:discussion}

\subsection{Summary of Results}

LVC parameter estimation found that the most likely scenario for GW190521 was an equal-mass ($q > 0.3$) BBH with both constituents falling inside the (pulsational) pair instability mass gap, and that the binary's effective spin was consistent with zero \cite{lvc_properties_GW190521}. This result was obtained under priors uniform in detector-frame BH masses and isotropic in constituent spins. Population-informed priors on the secondary mass led to the gap straddler interpretation of the LVC samples by \citet{gap_stradler_fishbach2020}, and priors uniform in the source frame total mass and inverse mass ratio led to the discovery by \citet{nitz2020_game_over} of solutions with higher likelihood at mass ratios far from unity. In Section~\ref{sec:posteriors} we presented posteriors sampled under the four different pairings of the mass and spin priors described in Section~\ref{sec:priors}. Based on the sensitivity of GW190521 source parameter inference to this choice between uninformative priors, it is difficult to confidently infer parameters without the strong assumption that the mass and spin priors describe the true merger population in the event's formation channel.

In order to say something about these different solution regions without assuming a specific mass and spin distribution, we consider the ensembles of samples across all four priors. At this point we stop speaking in terms of Bayesian evidence and think of the samples instead as a likelihood map, indicating which parameter space regions fit the data well. From this vantage point we cut the $q$--$\chieff$ space into quadrants and find that the only region producing a GW190521 solution with both masses outside the gap is the unequal-mass ($q < 0.3$) region with negative effective spin. The solution regions in this breakdown (see Equation~\eqref{eqn:p_defs_qchi}) agree on the sky position, but the characteristic distance and likelihood differ, with an increase in likelihood and a decrease in distance when moving toward more negative effective spin and mass ratio farther from unity (see Figure~\ref{fig:m12_lnL_DL_by_qchi_mode}).

We illustrate this competition between geometric prior volume and likelihood by computing the ratios of sensitive $VT(\pint)$ for the representative samples in Table~\ref{table:soln_mode_examples}, which tell us about the intrinsic luminosity of each solution region independent of the GW190521 data, and the ratios of $D_{\rm{com}}^3 \mathcal{L}$, which give a rough idea of the relative probabilities suggested by the data after accounting for the comoving volume effect but before mass and spin priors are introduced. The results in Table~\ref{table:soln_mode_examples} show that the equal-mass region with non-negative spin ($\pint^{E+}$) has a sensitive $VT$ larger than that of the unequal-mass region with negative effective spin ($\pint^{U-}$) by a factor of $\mathcal{O}(10)$. This geometric prior advantage is overcome by the characteristic likelihood in $D_{\rm{com}}^3 \mathcal{L}$, with the representative sample ratio favoring $\pint^{U-}$ over $\pint^{E+}$ by a factor of $\mathcal{O}(100)$.

Turning to a breakdown by number of BH masses in the gap of [45, 135] $\msun$, we confirm that the scenario with both masses outside the gap approximately corresponds to $\pint^{U-}$ and the double gap scenario maps onto both the positive and negative effective spin branches of the equal-mass peak. Repeating the calculations above with representative samples from the gap breakdown defined in Equation~\eqref{eqn:p_defs_ngap}, we see that the results in Table~\ref{table:gap_mode_examples} paint a similar picture to the $q$--$\chieff$ breakdown. The double gap region has a sensitive $VT$ advantage of $\mathcal{O}(10)$ over the region that avoids the mass gap, but the ratios of characteristic $D_{\rm{com}}^3 \mathcal{L}$ suggest that, after including the distance effect but before considering mass and spin priors, the data prefer to move both BHs outside the gap by a factor of $\mathcal{O}(100)$ over the double gap scenario.

\subsection{Astrophysical Implications}

These rough estimates from representative samples can only provide qualitative intuition about what to expect from posteriors under a prospective set of mass and spin priors. There is, however, a conclusion from the likelihood map that appears robust to priors and to reasonable changes in the bounds of the mass gap: parameters that fit the data well either place at least one BH inside the mass gap or avoid the gap with a mass ratio far from unity and a precessing primary spinning opposite the direction of the orbit. Though one might have hoped that moving both BHs out of the mass gap could help the case for an isolated binary evolution origin, we find that such solutions have spin orientations that are characteristic of dynamical formation \cite{field_binary_no_chieff_pop2020, need_dynamical_to_misalign_orbit_and_spin_Rodriguez2016, signatures_of_dynamical_mapelli2020}. Producing mergers with negative effective spin and precession without dynamical formation requires a combination of special circumstances, such as might be found in a triple or higher-multiple system where the Lidov-Kazai effect is aided by dynamical interactions with additional compact companions to create extreme spin-orbit misalignment \cite{lidov_kozai_neg_chieff_triples_Liu_2017}.

With at least one BH in the mass gap, the spins are consistent with standard binary co-evolution but explaining the masses requires a refined understanding of stellar collapse \cite{Chen_massgap_collapse_simulations_2014, yoshida_mass_gap_PPI_simulations, young_star_clusters_populating_mass_gap2020a, metallicity_effects_mapelli2017, metallicity_in_young_star_clusters2020a, low_metallicity_1gBHs_to85m_shell_interactions_mass_gap_Farrell2021, isolated_evolution_dredge-up_mass-gap_CostaMapelli2020, Tanikawa2020popIII, Liu2020PopIIINSC, stellar_merger_scenario_CE_redshift_RanzoJiang2020} and/or accretion \cite{imbh_from_globular_clusters2002, relativistic_accretion_formation_scenario_simulations2021, accretion_for_stellar_mass_case_Rice2021} to avoid calling upon dynamical channels where hierarchical mergers could contribute to observed merger rates \cite{young_star_clusters_populating_mass_gap2020a, young_star_clusters_heavy_remnants2021, cluster_hierarchical_metallicity_spin_mapelli2021, hierarchical_from_dynamical_in_any_star_cluster2020b, migration_traps_spins_rates_mckernan_ford2020a}. In dense stellar environments the total rates of dynamical and isolated mergers can be comparable \cite{star_cluster_dynamical_v_isolated_rates_mapelli2020}, but additional compact companions may be necessary to aid in (re)capturing a remnant and merging the higher-generation system on observable time scales \cite{hierarchical_7merger_scenario2020b, hierarchical_from_triples2021, hierarchical_multiples_2g_190521_190814_Liu2020a}. Analyzing these phenomena requires modelling remnant natal kicks, and time scale constraints cause hierarchical merger rate densities to be sensitive to the BH spin distribution \cite{rate_hierarchical_cluster_spinningLO_nospinHI_rodriguez2019}.

One population that has dynamical mechanisms for producing both mass gap mergers and IMBH mergers with negative effective spin is the disk of an active galactic nucleus (AGN) \cite{agn_accretion_disk_merger_population2020a, mass_gap_agn_bbh_mergers2021}. Following the announcement of an electromagnetic (EM) counterpart candidate from \agn \cite{em_counterpart_agn}, analyses of GW190521 have also evaluated the evidence of posteriors under priors that fix the source to the host AGN's sky location \cite{samples_conditioned_on_agn, nitz2020_game_over, new_scoop_waveform_model_comparison_xphm2021}, but there is not enough evidence to confidently associate the events \cite{em_confident_association_unlikely_fishbach2021}. Independent of this possible EM counterpart, one might consider the Bayesian evidence of GW190521 PE under mass and spin priors describing the BBH merger population in AGN disks. The distribution is believed to split into sub-populations in the bulk of the disk and in migration traps \cite{migration_traps_spins_rates_mckernan_ford2020a}. The mass and spin characteristics of each merger population can be estimated with simulations, although the results depend on the assumed distribution of BH masses, natal spins and kicks \cite{agn_bbh_population_chieff_q_simulation_mckernan_ford2019}. AGN are also predicted to be effective at producing eccentric BBH mergers \cite{agn_factories_for_eccentric2020b, eccentric_merger_rates_in_agn2021}, which might make up a significant fraction of the AGN disk mergers in the sensitive band of ground-based detectors \cite{detectable_eccentric_mergers_in_agn_capture2020b}.

What we can say without knowledge of the true population is that the source-frame total mass is consistently unimodal and centered near $\sim$150 $\msun$, meaning the claim that the remnant is an IMBH is robust. There is also a prior-independent lack of evidence to rule out precession. The data is fit best by solutions with mass ratio far from unity, where the mass gap is avoided but the effective spin is negative and the primary is precessing. However, even these apparently robust statements rely on the assumption of a quasi-circular BBH inspiral, and an eccentric merger interpretation of GW190521 was proposed by \citet{eccentricGW190521_gayathri2020a} based on the comparable fit to the data obtained from 325 NR simulations of eccentric mergers with generic spin and inverse mass ratio up to 7. Regardless of the true source properties, a full PE that includes the additional dimension of eccentricity would almost certainly find new regions of comparable likelihood considering the degeneracy between precession and extreme eccentricity for IMBH mergers \cite{heavy_eccentric_head-on_confused_for_precessing2020} and so few signal cycles in the sensitive band to resolve it. As more heavy events are detected, the indeterminacy we see for GW190521 might become a recurring theme. This would make our understanding of the observed population sensitive to the way we resolve these high-mass degeneracies. The most important developments moving forward will be broadening the sensitive band of detectors to access lower frequencies where IMBH inspiral signals fall, and improving waveform models by increasing the NR waveform catalog for calibrating efficient interpolants for generically spinning, eccentric binaries with higher-order harmonics over a wide range of mass ratios.

\section*{Acknowledgements}

The authors would like to thank Geraint Pratten for his assistance with using IMRPhenomXPHM in software development. SO acknowledges support from the National Science Foundation Graduate Research Fellowship Program under Grant No. DGE-2039656. Any opinions, findings, and conclusions or recommendations expressed in this material are those of the authors and do not necessarily reflect the views of the National Science Foundation. HSC gratefully acknowledges support from the Rubicon Fellowship awarded by the Netherlands Organisation for Scientific Research (NWO). LD acknowledges support from the Michael M. Garland startup research grant at the University of California, Berkeley. TV acknowledges support by the National Science Foundation under Grant No. 2012086. BZ is supported by a research grant from the Ruth and Herman Albert Scholarship Program for New Scientists. MZ is supported by NSF grants PHY-1820775 the Canadian Institute for Advanced Research (CIFAR) Program on Gravity and the Extreme Universe and the Simons Foundation Modern Inflationary Cosmology initiative.

\vskip 4pt

This research has made use of data, software and/or web tools obtained from the Gravitational Wave Open Science Center (\url{https://www.gw-openscience.org/}), a service of LIGO Laboratory, the LIGO Scientific Collaboration and the Virgo Collaboration. LIGO Laboratory and Advanced LIGO are funded by the United States National Science Foundation (NSF) as well as the Science and Technology Facilities Council (STFC) of the United Kingdom, the Max-Planck-Society (MPS), and the State of Niedersachsen/Germany for support of the construction of Advanced LIGO and construction and operation of the GEO600 detector. Additional support for Advanced LIGO was provided by the Australian Research Council. Virgo is funded, through the European Gravitational Observatory (EGO), by the French Centre National de Recherche Scientifique (CNRS), the Italian Istituto Nazionale di Fisica Nucleare (INFN) and the Dutch Nikhef, with contributions by institutions from Belgium, Germany, Greece, Hungary, Ireland, Japan, Monaco, Poland, Portugal, Spain.

\bibliographystyle{apsrev4-1-etal}
\bibliography{references}

\end{document}